\documentclass[aps,pra,reprint,superscriptaddress]{revtex4-2}

\usepackage{amsmath,amsfonts,amssymb}
\usepackage{xcolor,graphicx}
\usepackage{tikz}
\usepackage{braket}
\usepackage{dcolumn}
\usepackage{bm}
\usepackage{subfig}
\usepackage{float}
\usepackage[pdftex]{hyperref}
\hypersetup{
	colorlinks = true,
	linkcolor = blue,
	anchorcolor = blue,
	citecolor = red,
	filecolor = blue,
	urlcolor = blue}

\begin{document}

\title{Non-Hermitian transport in Glauber-Fock optical lattices}

\author{Ivan Bocanegra}
\email[e-mail: ]{ivanbocanegrag@gmail.com}
\affiliation{Instituto Nacional de Astrof\'isica \'Optica y Electr\'onica, Calle Luis Enrique Erro No.1, Santa Mar\'ia Tonanzintla, Puebla, 72840, M\'exico}
\author{H\'ector M. Moya-Cessa}
\affiliation{Instituto Nacional de Astrof\'isica \'Optica y Electr\'onica, Calle Luis Enrique Erro No.1, Santa Mar\'ia Tonanzintla, Puebla, 72840, M\'exico}

\date{\today}

\begin{abstract}
The effect of a non-unitary transformation on an initial Hermitian operator is studied. The initial (Hermitian) optical system is a Glauber-Fock optical lattice. The resulting non-Hermitian Hamiltonian models an anisotropic (Glauber-Fock) waveguide array of the Hatano-Nelson-type. Several cases are analyzed and exact analytical solutions for both the Hermitian and non-Hermitian Schr\"odinger problems are given, as they are simply connected. Indeed, such transformation can be regarded as a non-unitary Supersymmetric (SUSY) transformation and the resulting non-Hermitian Hamiltonian can be considered as representing an open system that interchanges energy with the environment.
\end{abstract}
\maketitle

\section{Introduction}
In the last decades, optical lattices (or waveguide arrays) have got major attention for presenting new (discrete) diffraction effects, in comparison with the corresponding (continuous) diffraction that appears in the bulk. Such discrete diffraction can be tailored by properly adjusting the configuration of the array \cite{Christodoulides2003,Yariv,Morandotti1999,Eisenberg2000,Pertsch2002}. Besides, optical lattices are suitable optical devices to both simulate and study a considerable number of quantum and classical effects, for instance squeezed and coherent states \cite{Leon-Montiel2010,Leon-Montiel2011,Perez-Leija2010,Keil2011,Villegas-Martinez2022}, Talbot effect \cite{Iwanow2005,Rai2008}, classical and quantum walks \cite{Perets2008}, to mention a few (see also \cite{Keil2012,Perez-Leija2012,Szameit2007,Bromberg2009,Rodriguez-Lara2011}), in relativistic and non-relativistic \cite{Longhi2010,Dreisow2009} schemes. Examples encompassing both linear and non-linear optical systems can be easily encountered (see for instance Ref.\cite{Christodoulides2003}). Moreover, optical lattices have straightforward use in the treatment of optical information. For example, see the application of optical lattices as converters of modes developed in Ref. \cite{Heinrich2014}.

One well-studied lattice is the so-called ``Glauber-Fock" optical lattice, characterized by a non-uniform distance between each pair of adjacent waveguides \cite{Perez-Leija2010,Keil2011,Keil2012,Perez-Leija2012}. For any array, the coupling $ g$ between adjacent waveguides depends on the distance $d$ between them, as $ g\propto e^{-d}\in\mathbb{R}$ (upper panel in Figure \ref{fig.coupled}). Therefore, in the Glauber-Fock lattice (lower panel in Figure \ref{fig.coupled}), where the waveguides get closer as the site $n$ of the waveguide increases, the coupling between the $n$-th and the $(n+1)$-th waveguide is proportional to $\sqrt{n+1}$. Naturally, a physical (realizable) system contains a finite number $N$ of waveguides, [see a) in the lower panel of Figure \ref{fig.coupled}], however as $N\to\infty$ the array may be considered effectively as semi-infinite [b), in the lower panel of Figure \ref{fig.coupled}]. There are other well-studied lattices, apart from Glauber-Fock, with non-uniform spacing between subsequent waveguides (the reader is referred to \cite{Perez-Leija2011,Perez-Leija2013,Perez-Leija2013b}), presenting intriguing transport characteristics as well.

More recently, waveguide arrays have as well been considered in the non-Hermitian regime. PT-symmetry plays an important role in this context \cite{Bender1998,Bender1999,Mostafazadeh2003}, presenting a plethora of new features, for instance invisibility \cite{Longhi2015,Regensburger2012}, phase transitions at exceptional points, power oscillations, double refraction \cite{Makris2008,Makris2010}, etc. In parallel, non-Hermitian arrays have also been considered (in minor proportion) under the Hatano-Nelson model \cite{Longhi2015,Yuce2022,Weidemann2020} (see also \cite{Liu2021,Liu2022,Bocanegra2023ES}). This in turn is just a (rather simple) non-unitary transformation of a conventional Schr\"odinger equation
\begin{equation}
    i\frac{\partial |\Psi(t)\rangle}{\partial t}=\left[\frac{\hat{p}^2}{2}+V({x})\right]|\Psi(t)\rangle.
\label{Schr}
\end{equation}
For  
\begin{equation}
|\Psi(t)\rangle=e^{ \gamma x} |\Phi(t)\rangle, \qquad  \gamma\in\mathbb R,
\label{transformation}
\end{equation}
the non-Hermitian Schr\"odinger type equation \cite{Hatano}
\begin{equation}
    i\frac{\partial |\Phi(t)\rangle}{\partial t}=\left[\frac{(\hat{p}+i \gamma)^2}{2}+V({x})\right]|\Phi(t)\rangle,
    \label{nSchr}
\end{equation}
is obtained. Indeed, non-unitary transformations produce non-Hermitian dynamics naturally \cite{Braulio}. In the optical context, a transformation of the type in (\ref{transformation}) produces an anisotropic (also called non-reciprocal) optical lattice;  due to the non-Hermiticity of the corresponding Hamiltonian, this might also be interpreted as an open system of waveguides (one interacting with the surroundings). It is worth to highlight that (\ref{transformation}) can be upturned, thus giving the solutions to the non-Hermitian equation (\ref{nSchr}) from the solutions of the Hermitian equation (\ref{Schr}), by means of a straightforward transformation.

Therefore, the non-Hermitian propagation in a Glauber-Fock optical lattice is studied in what follows, from the solution of the corresponding Hermitian system. The resulting non-Hermitian optical lattice then represents an anisotropic system of waveguides, for which exact solutions of the corresponding dynamical equations are easily obtained. At this point is worth to remark that such non-reciprocal systems of the Hatano-Nelson type can indeed be implemented in the laboratory (see Ref. \cite{Liu2022}).

The contents of the article are given in the following order: in section \ref{sec.equally}, the general frame regarding the Glauber-Fock optical lattice is established. In the same section, the mechanism for obtaining the non-Hermitian propagation departing from the Hermitian system is given. In sections \ref{sec.semi} and \ref{sec.finite}, respectively, the semi-infinite and finite cases are discussed, from the foundations developed in section \ref{sec.equally}. In section \ref{sec.su11} and \ref{sec.driven}, a pair of modifications of the Glauber-Fock lattice are addressed. Finally, the conclusions are given in section \ref{sec.conclusions}.
\section{Glauber-Fock waveguide lattice}
\label{sec.equally}

\begin{figure}[h]
    \centering
{\includegraphics[width=\linewidth]{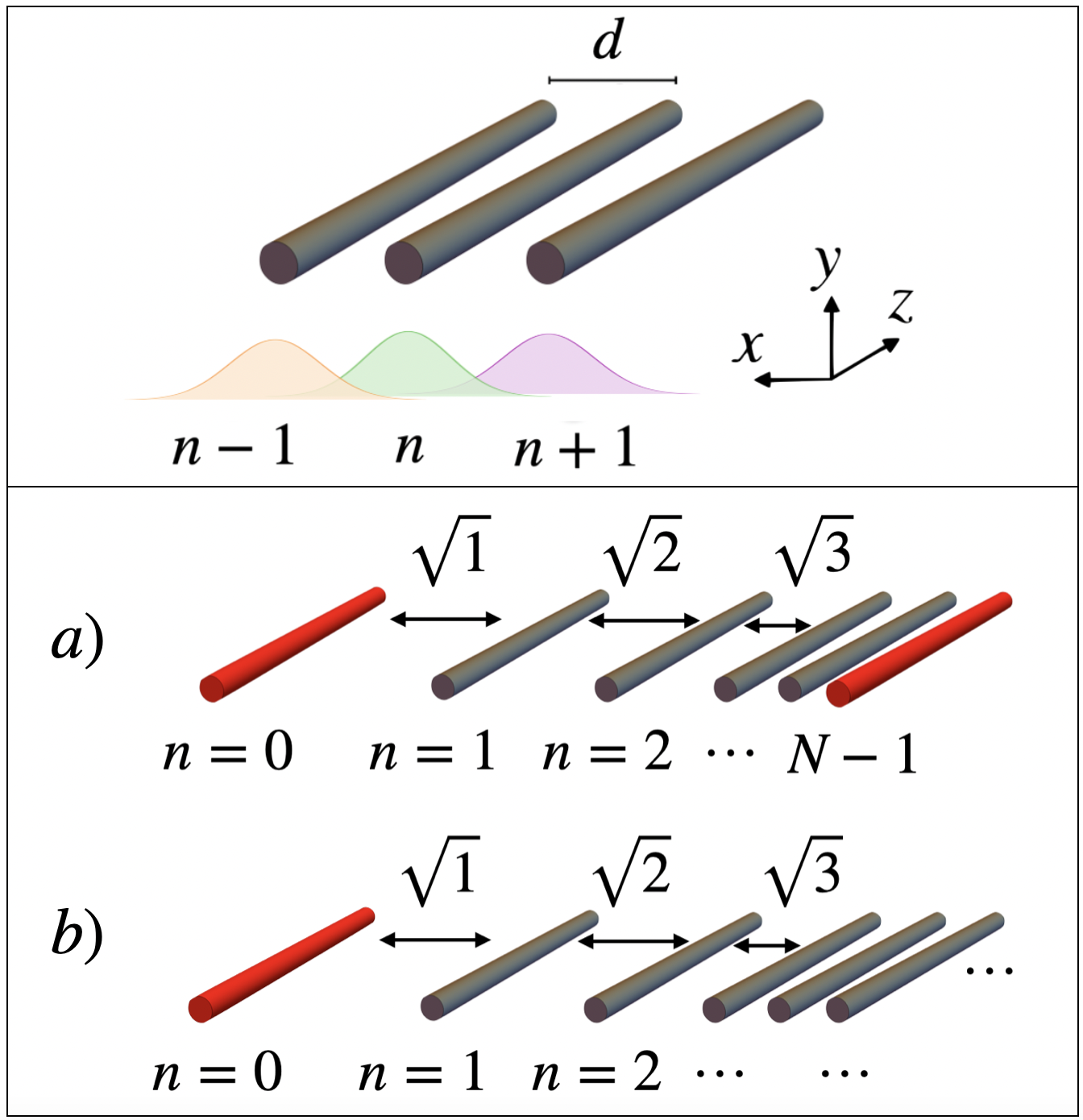}}
\caption{Upper panel: illustration of the coupling of modes between contiguous waveguides by means of the evanescent fields. The Gaussian distributions represent the modes of the electric field amplitude in the transversal direction $x$ and the coupling occurs where the Gaussian functions overlap.  In turn, such coupling can be tailored by adjusting the distance $d$ is between two neighboring waveguides. Lower panel: in the Glauber-Fock array, the waveguides get closer as the site $n$ of the waveguide increases, such that the coupling between the $n$-th and the $(n+1)$-th waveguide is proportional to $\sqrt{n+1}$. A distinction between a) the finite and b) the semi-infinite regimes is made. In a), we consider a number $N$ of waveguides, for example $n$ going from $0$ to $N-1$. Thus, two edge waveguides (highlighted in red color) are present, at $n=0$ and $n=N-1$. However, in b) we consider that $N\to\infty$, such that there exists a single edge waveguide (also in red), and the optical lattice is regarded as semi-infinite. }
	\label{fig.coupled}
\end{figure}

Quite generally, the amplitude $c_n(z)$, $n\in A\subseteq \mathbb Z$, of the electric field propagating in the $n$-th waveguide of a tight-binding waveguide array is coupled to the amplitudes $c_{n-1}(z)$ and $c_{n+1}(z)$ propagating in the contiguous waveguides \cite{Yariv} (upper panel in Figure \ref{fig.coupled}). Particularly, for the Glauber-Fock lattice (lower panel in Figure \ref{fig.coupled}), the propagation of the electric field amplitude $c_n(z)$ is ruled by
\begin{equation}
    i\dot c_n(z)+  g[\sqrt{n}c_{n-1}(z)+\sqrt{n+1}c_{n+1}(z)]=0,
\label{optical}
\end{equation}
where the dot stands for derivative with respect to the evolution parameter $z$, $ g=\kappa e^{-d}\in\mathbb{R}
$ is the coupling constant between the first two waveguides, with $\kappa$ a given real constant, and $d$ the corresponding distance between the first two sites. The equality (\ref{optical}) is related to an equation of the Schr\"odinger type  
\begin{equation}
    i\frac{\partial |\psi(z)\rangle}{\partial z} = H|\psi(z)\rangle,
\label{quantum}
\end{equation}
with
\begin{equation}
    H=- g(a^{{\dagger}}+a),
\label{Hamilt}
\end{equation}
the Hermitian Hamiltonian operator, namely $H=H^{\dagger}$, and where $a^\dagger$ and $a$ are raising and lowering operators, respectively \cite{Louisell}. In the semi-infinite case $a^\dagger$ and $a$ are simply the creation and annihilation operators of the harmonic oscillator. 
It is assumed that each $c_n(z)$ is associated to an abstract vector $|n\rangle\in\mathcal H$, with $\mathcal H$ a Hilbert space. Thus, (\ref{optical}) and (\ref{quantum}) are related by 
\begin{equation}
|\psi(z)\rangle=\sum_{n} c_n(z)|n\rangle.
\label{linear}
\end{equation}
Physically, the vector $|\psi(z)\rangle\in\mathcal H$ contains the information of the total electric field amplitude in the lattice for all the values of the propagation distance $z$. Then, $\displaystyle\sum_{n} |c_n(z)|^2=1$ is a suitable choice for normalization.

The electric field amplitudes are given by
\begin{equation}
c_{j}(z)=\langle j|\psi(z)\rangle,\qquad j\in A,
\label{ceka}
\end{equation}
where the set $A$ is going to be defined according to the dimension of the corresponding Hilbert space. The same is true for the ladder operators $a^\dagger$ and $a$. In what follows, a Hilbert space $\mathcal H$ of generic dimension is considered, in order to establish the methodology to generate the non-Hermitian transport in the Glauber-Fock lattice. The precise dimensions of the Hilbert space for each case (finite and semi-infinite) will be specified in the corresponding sections.
\subsection{Non-Hermitian Glauber-Fock lattice}

The solution of equation (\ref{quantum}) is proposed as
\begin{equation}
|\psi(z)\rangle=e^{-\gamma \hat n} |\phi(z)\rangle,
\qquad \gamma\in\mathbb R,
\label{psiphi}
\end{equation}
where $\hat n |j\rangle=j|j\rangle$, $j\in A$. Then, the non-Hermitian (non-conservative) problem
\begin{equation}
    i\frac{\partial |\phi(z)\rangle}{\partial z} = \tilde H|\phi(z)\rangle,
\label{nHermitian}
\end{equation}
is obtained, where
\begin{equation}
    \tilde H=e^{\gamma\hat n}He^{-\gamma\hat n}=-(k_1a^{\dagger}
+k_2a),
\label{htilde}
\end{equation}
 is clearly non-Hermitian, namely $\tilde H^\dagger\neq \tilde H$, with $k_1=  g e^\gamma$ and $k_2= g e^{-\gamma}$. 
 The last equality in (\ref{htilde}) comes from the commutators $[\hat n,a]=-a$ and $[\hat n,a^\dagger]=a^\dagger$ . Our main concern is then the non-Hermitian system (\ref{nHermitian})-(\ref{htilde}).
 
In the specific case of the semi-infinite Glauber-Fock lattice, the Hamiltonian (\ref{htilde}) is of the type of that studied in Ref. \cite{Yuce2022}. Therefore, the Hatano-Nelson problem (\ref{nHermitian})-(\ref{htilde}) is by itself pretty interesting. The reader interested in the implementation, as well as the study, of systems of the type of Hatano-Nelson in equally-spaced optical lattices is referred to \cite{Longhi2015,Weidemann2020,Liu2021,Liu2022,Bocanegra2023ES}. Here we analyze (\ref{nHermitian})-(\ref{htilde}) for several systems with open and closed boundary conditions, and give the corresponding exact analytical solutions, not present in the available literature. 

The equation (\ref{nHermitian}) is connected to the equation
\begin{equation}
    i\dot d_n(z)+k_1 \sqrt{n} d_{n-1}(z)+k_2 \sqrt{n+1}d_{n+1}(z)=0,
\label{neoptical}
\end{equation}
via
\begin{equation}
|\phi(z)\rangle=\sum_{n} d_n(z)|n\rangle,
\label{dlinear}
\end{equation}
with 
\begin{equation}
    d_k(z)=\langle k
|\phi(z)\rangle=e^{\gamma k}c_k(z), \qquad
k\in A,
\label{dn}
\end{equation} standing for the electric field amplitude in the $k$-th waveguide of the non-Hermitian (non-conservative) lattice. 

\begin{figure}[h]
    \centering
    {\includegraphics[width=\linewidth]{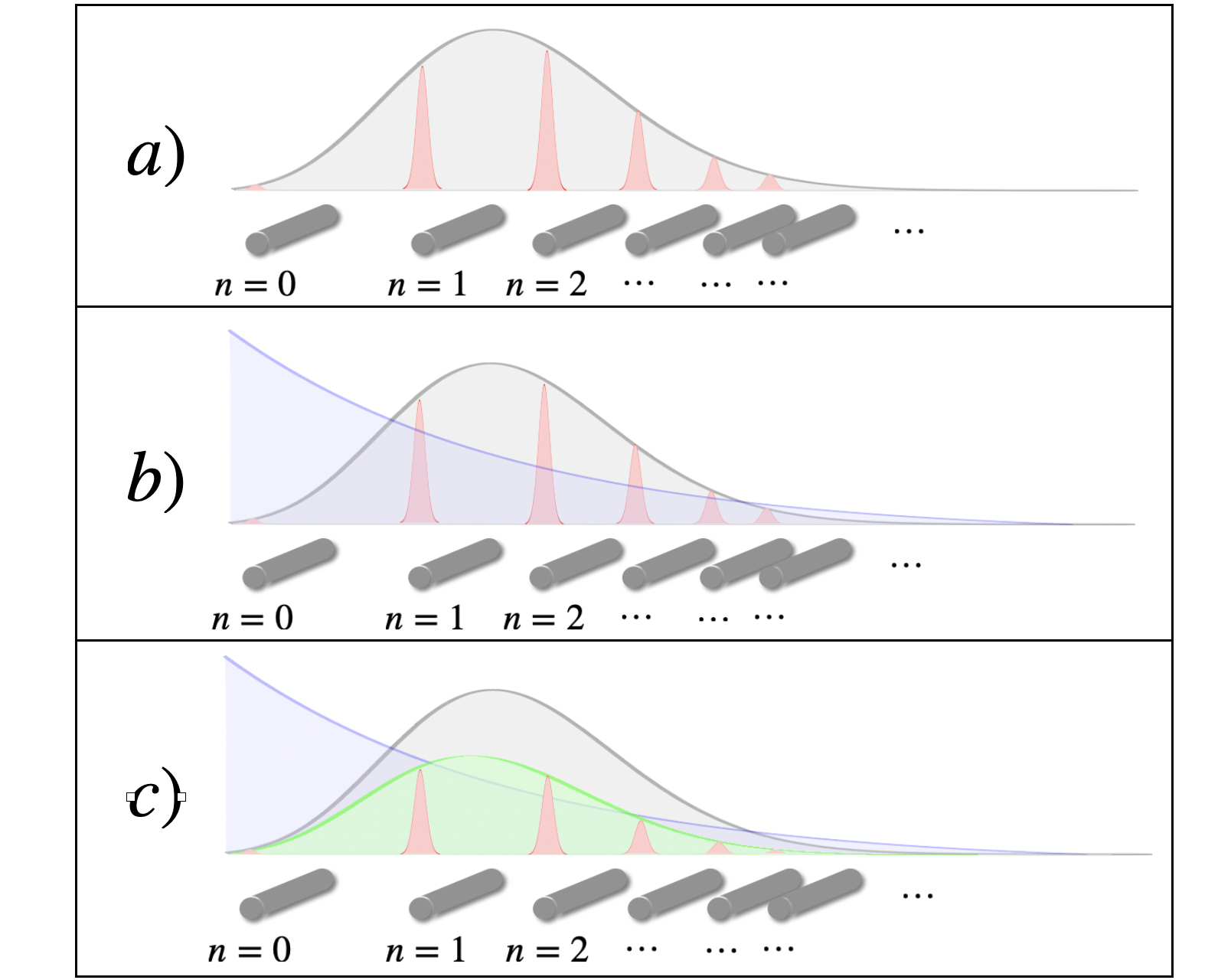}}
\caption{Illustrative action of the non-unitary transformation (\ref{phipsi}) on a state $|\psi(0)\rangle$ with coefficients $c_n(0)$ initially distributed according to a Poissonian function, for a Glauber-Fock optical lattice in the semi-infinite regime. In a), the red Gaussian distributions, with heights given by $c_n(0)$ for all $n$, represent the modes in each waveguide. The Poisson distribution envelope is represented in gray color. As before (upper panel in Figure \ref{fig.coupled}), the horizontal axis coincides with the coordinate $x$ of the frame of reference. In b), the non-unitary transformation (\ref{phipsi}) is schematically pictured by a blue decreasing exponential ($\gamma<0$) operation $e^{\gamma \hat n}$ on the initial state $|\psi(0)\rangle$. The output state $|\phi(0)\rangle=e^{\gamma \hat n}|\psi(0)\rangle$ is shown in c). According to (\ref{dn}), the
envelope of the corresponding output coefficients $d_n(0)=e^{\gamma n}c_n(0)$ is given in green. For $\gamma>0$, the operation $e^{\gamma \hat n}$ amplifies the state $|\psi(0)\rangle$ instead.}
	\label{fig.transformation}
\end{figure}

The solution $|\phi(z)\rangle$ of (\ref{nHermitian}) can be obtained from the solution $|\psi(z)\rangle$ of the associated Hermitian problem (\ref{quantum}), in the case of general $k_1$ and $k_2$ (cf. \cite{Yuce2022}), as
\begin{equation}
  |\phi(z)\rangle=  \left(\frac{k_1}{k_2}\right)^{\displaystyle\frac{\hat n}{2}}|\psi(z)\rangle .
\label{gphipsi}
\end{equation}
Then, for $k_1$ and $k_2$ given below (\ref{htilde}), the non-Hermitian propagation is straightforwardly obtained as
\begin{equation}
 |\phi(z)\rangle=  e^{\gamma\hat n}|\psi(z)\rangle .
\label{phipsi}
\end{equation}
In Figure \ref{fig.transformation}, the outcome of the non-unitary transformation (\ref{phipsi}) on an initial state $|\psi(0)\rangle$ is illustratively shown. It produces an amplification ($\gamma>0$) or attenuation ($\gamma<0$) of the distribution of the electric field. Both effects can be awarded to an external process. The output state
$|\phi(0)\rangle=e^{\gamma \hat n}|\psi(0)\rangle$, and in general $|\phi(z)\rangle$, is not normalized as the electric field in the array is either being someway absorbed or the array is being provided with energy from the exterior, depending to the sign of $\gamma$ in (\ref{phipsi}). Indeed, such an amplification or attenuation can be easily implemented in the laboratory \cite{Ruter2010}.

\begin{figure}[h]
    \centering
{\includegraphics[width=\linewidth]{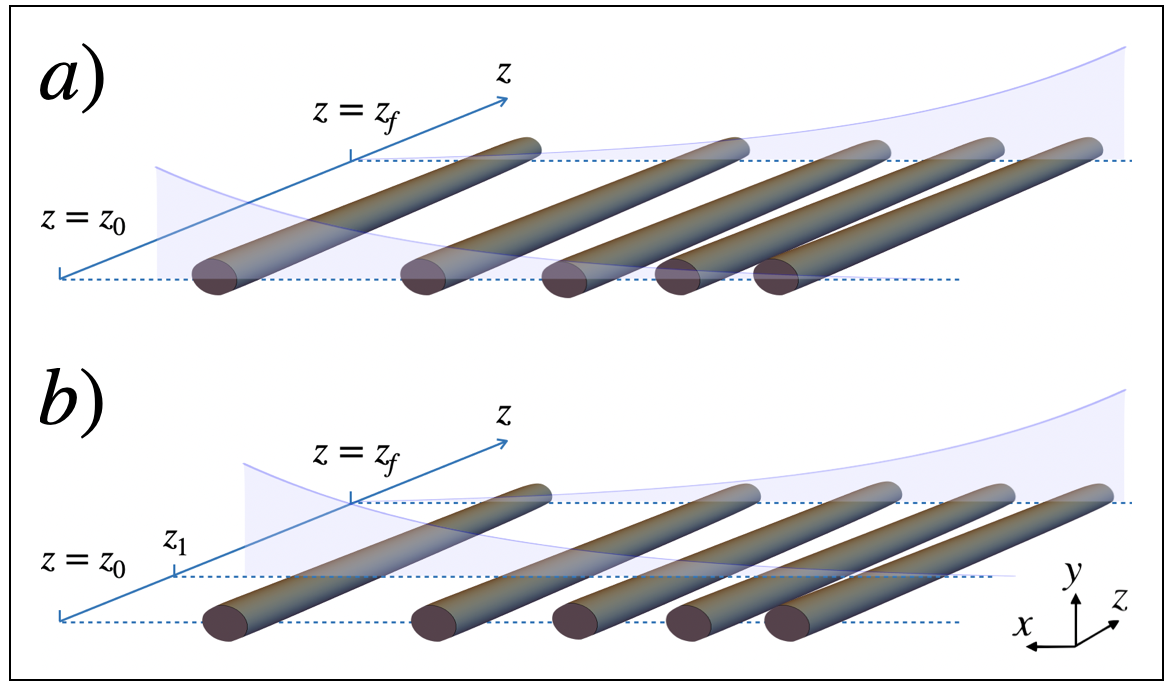}}
\caption{Effect of the transformation (\ref{phipsi}) in the transport along $z$. a) The transformation (\ref{phipsi}) is applied on $|\psi(z_0)\rangle$ at $z=z_0$. Afterwards, the inverse transformation (\ref{psiphi}) is performed, at $z=z_f$, on $|\phi(z_f)\rangle$, this take the state of the system back to $|\psi(z_f)\rangle$. Certainly, the state $|\phi(z)\rangle$ represents the non-Hermitian propagation in the interval $z_0\leq z<z_f$. In b), the transformation (\ref{phipsi}) is illustratively performed at $z_1>z_0$ and, once again, at $z_f$ the inverse transformation (\ref{psiphi}) is performed. In the interval $z_1\leq z< z_f$, the state $|\phi(z)\rangle$ represents non-Hermitian transport, while for $z<z_1$ and $z\geq z_f$ Hermitian transport is obtained. Therefore, intervals of non-Hermitian and Hermitian propagation can be alternated at will.}
	\label{fig.transport}
\end{figure}

Besides, in Figure \ref{fig.transport} it is sketched the way to implement the transformation (\ref{phipsi}) to get the corresponding non-Hermitian transport. Also it is shown the application of the inverse transformation (\ref{psiphi}), in order to recover the Hermitian propagation. In addition, due to the fact that both transformations (\ref{phipsi}) and (\ref{psiphi}) can be applied at any propagation distance $z$, alternating intervals of non-Hermitian and Hermitian transport can be readily implemented and studied.
\\ \\
\textit{Non-unitary transformation as a Supersymmetric transformation}. Indeed, the non-unitary transformation (\ref{phipsi}), can be regarded as a Supersymmetric (SUSY) transformation \cite{Cooper1995,Bocanegra2022,Bocanegra2023}, as the Hamiltonians $H$ and $\tilde H$ given in (\ref{Hamilt}) and (\ref{htilde}), respectively, are connected by the operator $T=e^{\gamma \hat n}$, as
\begin{equation}
    T H=\tilde H T. 
\label{inter}
\end{equation}
As we shall see in Section \ref{sec.finite}, in the stationary regime, the Hamiltonians $H$ and $\tilde H$ are isospectral, as expected from relation (\ref{inter}).
\section{Semi-infinite lattice}
\label{sec.semi}
In the case of the semi-infinite optical lattice the annihilation (creation) $a$ ($a^\dagger$) operator, together with the number operator $\hat n$ can be cast in the form 
    \begin{equation}
    a=\sum_{k=0}^{\infty} \sqrt{k+1}|k\rangle\langle k+1|,
    \end{equation}
    \begin{equation}a^{{\dagger}}=\sum_{k=0}^{\infty}\sqrt{k+1}|k+1\rangle\langle k|,
    \end{equation}
    \begin{equation} \hat{n}=\sum_{k=0}^{\infty}k |k\rangle\langle k|.
    \end{equation}
As usual, the commutation relations
    \begin{equation}[a,a^\dagger]=\mathbb I,
    \end{equation}
    \begin{equation}
    [\hat n,a]=-a,
    \label{commutation1}
    \end{equation}
    \begin{equation}
    [\hat n,a^\dagger]=a^\dagger,
    \label{commutation2}
    \end{equation}

hold. In turn,
(\ref{optical}) and (\ref{quantum}) are connected by
\begin{equation}
    |\psi(z)\rangle=\sum_{k=0}^\infty c_k(z)|k\rangle,
\label{combination}
\end{equation}
with $\left\{|k\rangle\right\}_{k=0,1,\dots}$ the Fock states. 

\begin{figure}[h]
    \centering
    {\includegraphics[width=\linewidth]{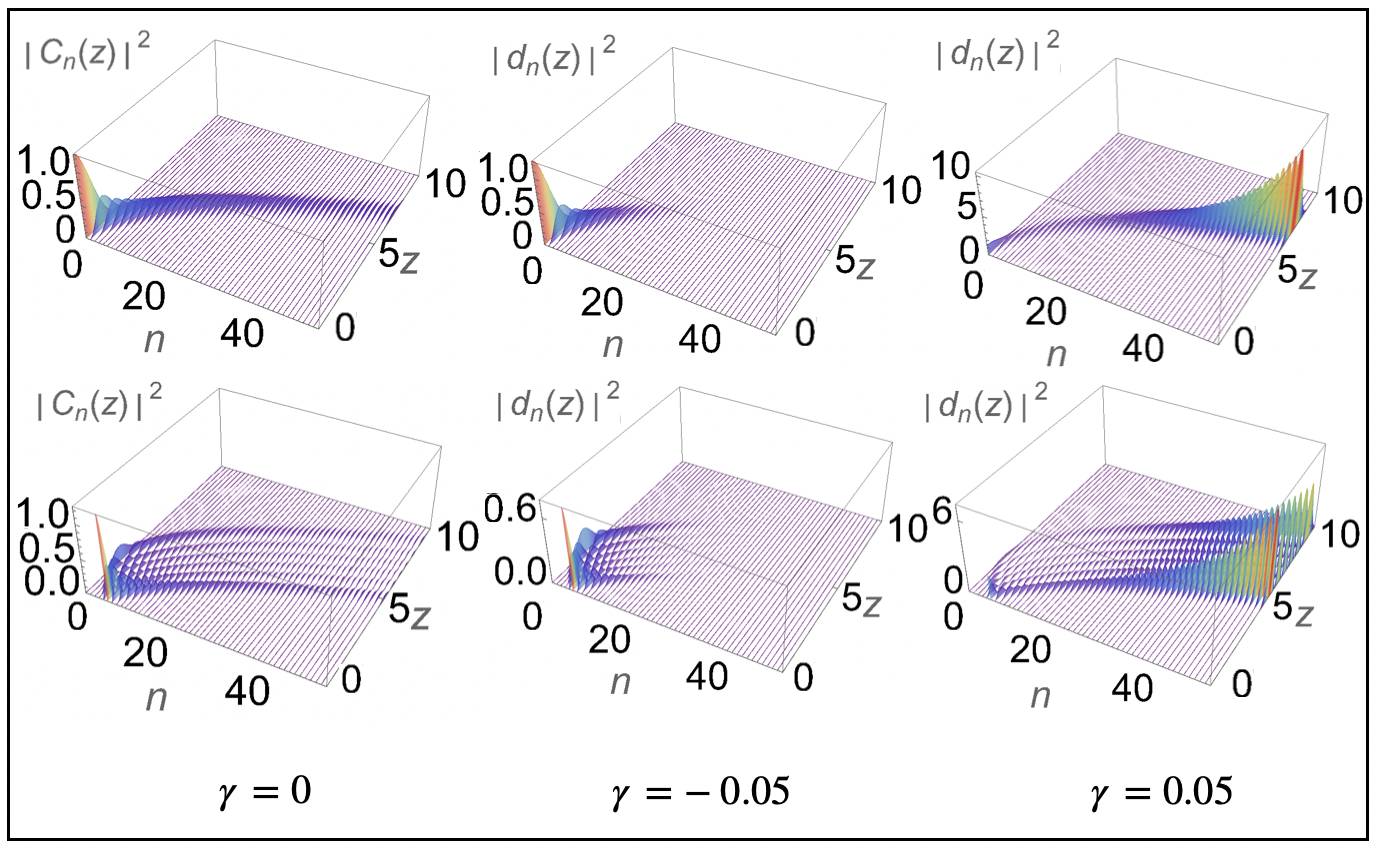}}
\caption{Non-Hermitian propagation of the intensity $|d_n(z)|^2$ in the semi-infinite Glauber-Fock optical lattice, as given by (\ref{dn1}), in the case of a single waveguide (situated at a given $m$) excited at $z=0$, and its comparison with the Hermitian case. In all the figures the propagation distance $z$ is measured in units of $g$. Besides, the parameter $g = 1$ has been set. In the upper row the waveguide at the edge $m=0$ is excited at $z=0$, in the Hermitian ($\gamma=0$, upper left) and non-Hermitian  ($\gamma\neq 0$, upper middle and upper right) scenarios. The transport is neither amplified nor attenuated in the Hermitian system (upper left). In turn, when $\gamma = -0.05$ ($\gamma = 0.05$), exponential attenuation (amplification) is shown (see also the vertical scales). This means that the array behaves as an open system, as it either looses electromagnetic power, or it is being provided with it from the exterior. In the lower row the propagation for the case of a waveguide in the bulk (for instance at site $m=5$) being excited at $z=0$ is shown. It can be seen that reflection occurs at the edge waveguide $m=0$. Again, there is neither amplification nor attenuation in the Hermitian system ($\gamma=0$, lower left). However, for $\gamma=-0.05$ ($\gamma=0.05$) the transformation (\ref{phipsi}) produces an attenuation (amplification) in the field as $n\to\infty$. It is worth to note that the number of maxima in the electromagnetic field distribution increases with $m$, due to the reflection at the boundary.}
	\label{fig.semiinfinite1}
\end{figure}

\begin{figure}[h]
    \centering
    {\includegraphics[width=\linewidth]{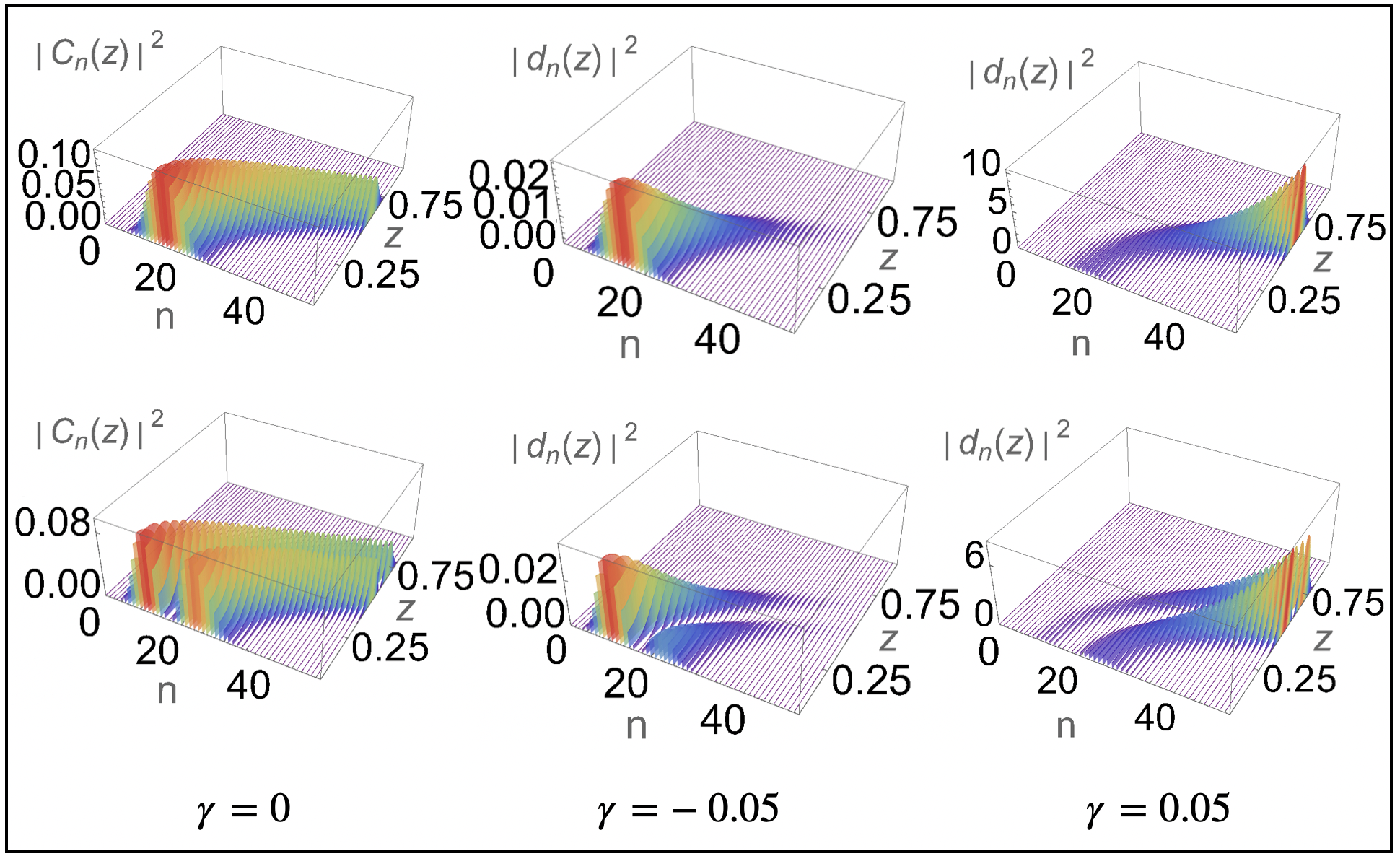}}
\caption{Non-Hermitian propagation of the intensity $|d_n(z)|^2$ in the semi-infinite Glauber-Fock optical lattice, as given by (\ref{displaced}), in a semi-infinite Glauber-Fock waveguide array when it is excited at $z=0$ with a distribution (a \textit{generalized coherent state}). The parameters are those of Figure \ref{fig.semiinfinite1} and $\chi=4$. For $m=0$ (upper row), the distribution injected at $z=0$ is Poissonian with mean equal to $16$. As in the case of the impulse (Figure \ref{fig.semiinfinite1}), the distribution follows a curved trajectory towards $n\to\infty$, what can be particularly appreciated in the Hermitian case $\gamma=0$ in the upper left panel (that in turn closely resembles the corresponding panel in Figure \ref{fig.semiinfinite1}). The non-Hermitian cases $\gamma\neq 0$ (upper middle and upper right) resemble pretty much the corresponding images in Figure \ref{fig.semiinfinite1} as well. This is, the transformation (\ref{phipsi}) attenuates or amplifies the electromagnetic field as $n\to\infty$, without affecting the trajectory. Such attenuation or amplification can be easily adjusted with the parameter $\gamma$. For $m=1$ (lower row) we have two lobes or maxima in the initial distribution at $z=0$, see for instance the Hermitian case $\gamma=0$ in the lower left panel.  Again, this resembles the case of the impulse (lower row in Figure \ref{fig.semiinfinite1}), as the number of maxima increases with $m$. The lobes follow a curved trajectory towards $n\to\infty$, as in the case $m=0$ (upper row). In turn, in the non-Hermitian cases $\gamma\neq 0$ (lower middle and lower right) the relative heights of such lobes can be tailored by adjusting the non-Hermitian parameter $\gamma$, without affecting its curved path. Once more this resembles the response to the impulse, as the effect of transformation (\ref{phipsi}) in the lower middle and lower right panels is quite similar to the one observed in the corresponding panels of Figure \ref{fig.semiinfinite1}. }
\label{fig.semiinfinite2}
\end{figure}

From (\ref{quantum}), we have
\begin{equation}
    |\psi(z)\rangle=e^{i  g z(a^\dagger+a)}|\psi(0)\rangle,
\label{DSG}
\end{equation}
with $|\psi(0)\rangle$ an arbitrary initial condition. We can identify the exponential operator in (\ref{DSG}) as $ D(i g z)$, where 
\begin{equation}
D(\xi)=e^{\xi a^\dagger-\xi^\star a}=e^{-\frac{1}{2}|\xi|^2}e^{\xi a^\dagger}e^{-\xi^\star a},\qquad \xi\in\mathbb C,
\label{Glaub}
\end{equation}
is the Glauber displacement operator \cite{Louisell}. The $*$ stands for complex conjugation.
\\ \\
\textit{Response to the impulse.}
We can set the initial condition $|\psi(0)\rangle$ in (\ref{DSG}) in such a way that at $z=0$ a single waveguide is excited, the $m$-th waveguide for example, this is $|\psi(0)\rangle=|m\rangle$, $m=0,1,\dots
$ In turn, the electric field $c_n(z)$ at the $n$-th waveguide is given by $\langle n|D(i g z)|m\rangle$. In general,
\begin{equation} 
\langle n| D(\xi)|m\rangle=e^{-\frac{1}{2}|\xi|^2}\xi
^{n-m}\sqrt{\frac{m!}{n!}}L_m^{n-m}(|\xi|^2),
\label{Saty}
\end{equation}
with $L_k^\ell$ the associated Laguerre polynomials of order $k$. Thus,
\begin{equation} 
c_n(z)=e^{-\frac{1}{2} g^2z^2}(i g z)^{n-m}\sqrt{\frac{m!}{n!}}L_m^{n-m}( g^2z^2).
\end{equation}
Such that in the non-conservative lattice the amplitudes of the electric field are directly
\begin{equation}
    d_n(z)=e^{\gamma n}e^{-\frac{1}{2} g^2z^2}(i g z)^{n-m}\sqrt{\frac{m!}{n!}}L_m^{n-m}( g^2z^2).
\label{dn1}
\end{equation}
For specific values of the involved parameters, the intensities $|d_n(z)|^2$ are shown in Figure \ref{fig.semiinfinite1}. The non-Hermitian transport given by (\ref{dn1}) can be compared with the corresponding Hermitian scenario $\gamma=0$. The upper row shows the response of the system when the edge waveguide $m=0$ is excited at $z=0$. In the Hermitian case $\gamma=0$ (upper left) no attenuation or amplification occurs. In turn, the field attenuates while propagating when $\gamma=-0.05$ (upper middle). Actually, as in any physical propagation phenomena there always exists loss because of the environment's interaction, the $\gamma<0$ case serves to model real transport phenomena. On the contrary, amplification is shown when $\gamma=0.05$ (upper right). Then, the case $\gamma>0$ serves to model external providing of electromagnetic field, for example, when losses are to be reduced through external feeding of electromagnetic power. In addition, the lower row shows the propagation obtained when the initially excited waveguide is in the bulk, for example at $m=5$. In such a case, reflection at the boundary $m=0$ is seen to occur: this can be clearly observed in the Hermitian regime (lower left). In the cases $\gamma=-0.05$ (lower middle) and $\gamma=0.05$ (lower right) it can be appreciated, respectively, attenuation and amplification as $n\to\infty$. It is worth to remark that the number of maxima of the electromagnetic field increases with $m$, due to the reflection at the boundary, and is equal to $m+1$. In the lower row ($m=5$) of Figure \ref{fig.semiinfinite1} we have $6$ maxima.
\\
\\
\textit{Coherent states as initial condition.}
The initial condition $|\psi(0)\rangle$ in (\ref{DSG}) is 
now chosen as $|\psi(0)\rangle=|m,\chi\rangle$, where $|m,\chi \rangle:=D(\chi)|m\rangle$ is a generalized coherent state (displaced number state), with $|m\rangle$ a Fock state and $D(\chi)$, $\chi\in\mathbb C$, given by (\ref{Glaub}). For $m=0$, the generalized coherent state $|m,\chi \rangle$ reduces to the conventional coherent state $|0,\chi\rangle\equiv|\chi\rangle$. Then, $c_n(z)=\langle n|D(i g z)|m,\chi\rangle$ is found to be
\begin{equation}
c_n(z)=e^{i g z \chi_R}e^{-\frac{1}{2}|\Omega |^2}\Omega^{n-m}\sqrt{\frac{m!}{n!}} L_m^{n-m}(|\Omega |^2),
\end{equation}
with $\Omega=\Omega( g,\chi,z):=\chi+i g z$, $\chi_R$ denoting the real part of $\chi$, and where (\ref{Saty}) has been used, together with the property
\begin{equation}
D( g)D(\beta)=D( g+\beta)e^{iIm( g\beta^*)}.
\end{equation}
The fields in the non-conservative system are just
\begin{equation}
d_n(z)=e^{\gamma n}e^{i g z \chi_R}e^{-\frac{1}{2}|\Omega |^2}\Omega^{n-m}\sqrt{\frac{m!}{n!}} L_m^{n-m}(|\Omega |^2).
\label{displaced}
\end{equation}
Figure \ref{fig.semiinfinite2} shows the transport along $z$ of the electromagnetic field in the non-conservative system of waveguides and its comparison with the Hermitian case ($\gamma=0$), according to (\ref{displaced}), for specific values of the parameters. Indeed, the propagation somehow resembles that of the impulse (Figure \ref{fig.semiinfinite1}). As in Figure \ref{fig.semiinfinite1}, the increasing of $m$ entails an increasing in the number of maxima. In this case in the maxima of the distribution. Also as in the case of the impulse, the transformation (\ref{phipsi}) allows to tailor the relative height of such lobes or maxima without affecting the curved trajectory towards $n\to\infty$.
\section{Finite array}
\label{sec.finite}
When the array is finite, the operators $a$, $a^\dagger$ in (\ref{Hamilt}), and $\hat n$, are given as
    \begin{equation}
    a_N=\sum_{k=0}^{N-1} \sqrt{k+1}|k\rangle\langle k+1|, 
    \label{foperators1}
    \end{equation}
    \begin{equation}
    a_N^{{\dagger}}=\sum_{k=0}^{N-1} \sqrt{k+1}|k+1\rangle\langle k|, 
    \label{foperators2}
   \end{equation}
    \begin{equation} \hat{n}_{N}=\sum_{k=0}^{N-1}k |k\rangle\langle k|.
    \label{foperators3}
    \end{equation}
The corresponding commutators are
    \begin{equation}[a_N,a_N^\dagger]=\mathbb I_N- N|N\rangle\langle N|,
    \label{conm2a}
    \end{equation}
    \begin{equation}
    [\hat n_N,a_N]=-a_N,
    \label{conm2b}
    \end{equation}
    \begin{equation}
    [\hat n_N,a_N^\dagger]=a_N^\dagger,
    \label{conm2c}
    \end{equation}
with $\mathbb I_N$ the identity in the $N$-dimensional space.
Due to the commutation relation (\ref{conm2a}), it is not possible to express the evolution operator in the form (\ref{Glaub}), as in the semi-infinite case. So in this section we follow a different procedure. Nonetheless, as the relations (\ref{conm2b}) and (\ref{conm2c}) have exactly the same form of relations (\ref{commutation1}) and (\ref{commutation2}), respectively, we still can construct the solution corresponding to the non-Hermitian system in terms of the solution of the Hermitian one. Thus, along this section we only call $a$, $a^\dagger$ and $\hat n$, respectively, the operators defined in (\ref{foperators1})-(\ref{foperators3}).

The equations
(\ref{optical}) and (\ref{quantum}) are related through 
\begin{equation}
    |\psi(z)\rangle=\sum_{k=0}^{N-1} c_k(z)|k\rangle,
\end{equation}
where $|k\rangle\in\mathcal H, k=0,1,\dots,N-1$. As there exist two edge (or ending) waveguides, the condition $c_\ell=0, \ell\geq N$, must be added. Then, along with (\ref{optical}) we have 
\begin{equation}
i\dot c_{N-1}(z)+  g\sqrt{N-1} c_{N-2}(z)=0,
\label{opticalfinite}
\end{equation}
and
\begin{equation}
i\dot d_{N-1}(z)+k_1 \sqrt{N-1}d_{N-2}(z)=0,
\end{equation}
along with (\ref{neoptical}).

The operator Hamiltonian  in (\ref{Hamilt}), for $a$ and $a^\dagger$ in (\ref{foperators1}) and (\ref{foperators2}), then acquires the form of a tri-diagonal square matrix of dimension $N$ (see  \cite{Rodriguez-Lara2011} and references therein)
\begin{equation}
    H= - g\left(\begin{array}{cccccc}
        0 & \sqrt{1} & 0  & \cdots & 0\\
        \sqrt{1} & 0 & \sqrt{2}  &\cdots & 0\\
        0 & \sqrt{2} & 0  &\cdots & 0\\
        \vdots & \vdots & \vdots & \ddots & 
        \vdots\\
        0 & 0 & \cdots & 0 & \sqrt{N-1}\\
        0 & 0 & \cdots & \sqrt{N-1} & 0\\
    \end{array}\right).
    \label{Hfinite}
\end{equation}
 Now, a (similarity) transformation is performed on $H$ as \begin{equation}
 \Lambda = S^{-1}MS, \qquad M\equiv-H,
 \label{similarity}
 \end{equation}
 where $S$ is an orthonormal matrix such that its columns are the (normalized) eigenvectors $|\psi_j\rangle$ of $M$, i.e.
 \begin{equation}
      M |\psi_j\rangle=\lambda_j |\psi_j\rangle, \qquad j=0, 1, \dots,N-1.
\label{eva}
 \end{equation}

The set of eigenvalues $\left\{\lambda_j\right\}$ is obtained from the usual condition
$D_N = 0$,
where $D_N\equiv$ det$(M-\lambda \mathbb I_N)$. Besides, as $H$ is tri-diagonal, $D_N$ is given by the recurrence relation
    \begin{equation}
  D_0=1,  
  \label{recurrence1}
    \end{equation}
    \begin{equation}
    D_1= \lambda, 
    \label{recurrence2}
    \end{equation}
    \begin{equation}
    D_s=\lambda D_{s-1}-D_{s-2}, \qquad s=2,3,\dots,N.
    \label{recurrence3}
    \end{equation}
In turn, the normalized eigenvectors of $H$ are
\begin{equation}
    |\psi_j\rangle=\left(\sum_{k=0}^{N-1} \left[D_k(\lambda_j)\right]^2\right)^{-1/2}\left(\begin{array}{c}
         D_0(\lambda_j) \\
         D_1(\lambda_j)\\
         \vdots\\ 
         D_{N-1}(\lambda_j)
          
    \end{array}\right).
\end{equation}
The relations (\ref{recurrence1})-(\ref{recurrence3}) turn out to be those of the Hermite polynomials \cite{Abramowitz}. Therefore, the condition giving the set of eigenvalues $\left\{\lambda_j\right\}$ of $M$ is 
\begin{equation}
    D_N(\lambda)=H_N\left(\frac{\lambda}{\sqrt{2}}\right)=0,
\end{equation}
with $H_k(x)$ the Hermite polynomial of order $k$.
The normalized eigenvectors are then
\begin{equation}
   |\psi_j\rangle =\left(\sum_{k=0}^{N-1} V_{j,k}^2\right)^{-1/2}\left(\begin{array}{c}
         V_{j,0} \\
         V_{j,1}\\
         \vdots\\ 
         V_{j,N-1}          
    \end{array}\right),
\end{equation}
\begin{equation}
V_{j,k}=\frac{1}{\sqrt{2^k k!}}H_k\left(\frac{\lambda_j}{\sqrt{2}}\right).
\end{equation}

\begin{figure}[h]
    \centering
    {\includegraphics[width=\linewidth]{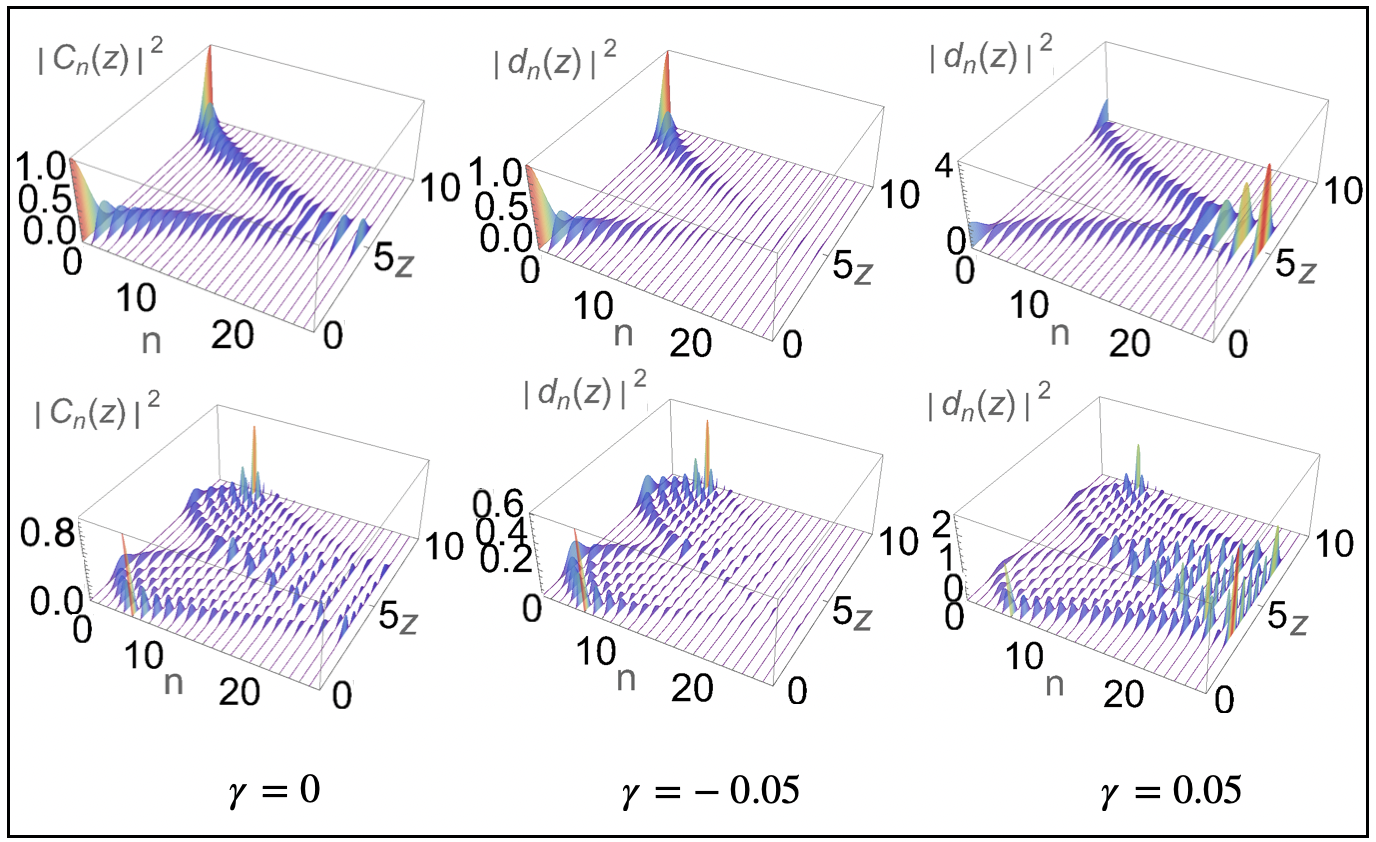}}
\caption{Non-Hermitian propagation of the intensity $|d_n(z)|^2$ in the semi-infinite Glauber-Fock optical lattice, as given by (\ref{finited}), in a finite Glauber-Fock optical lattices formed of $N=25$ waveguides for initial exciting of one single waveguide at the site $m$. In all the figures the longitudinal distance $z$ is measured in units of the coupling constant $g$, while the parameter $g= 1$ was set. Upper row: the end waveguide $m=0$ is excited at $z=0$. Pretty close to the corresponding row of Figure \ref{fig.semiinfinite1}, in the Hermitian regime $\gamma=0$ (upper left panel) there is no attenuation or amplification of the field. In turn, in the non-Hermitian case $\gamma=-0.05$ ($\gamma=0.05$) in the upper middle (upper right) panel, an attenuated (amplified) propagation can be appreciated. Such attenuation (amplification) is maximal at the opposite edge waveguide ($n=24$) and can be tailored by manipulating the parameter $\gamma$. In the lower row the waveguide situated at $m=5$ is excited at $z=0$. In this case there exist reflections at both end waveguides of the optical lattice (compare with the corresponding row in Figure \ref{fig.semiinfinite1}).}
	\label{fig.finite1}
\end{figure}

At this point it is worth to mention that $H$ and $M$ share common eigenvalues and its eigenvectors coincide up to a constant phase factor. Through reverting (\ref{similarity}) and replacing into $|\psi(z)\rangle=e^{-iHz}|\psi(0)\rangle$, it is obtained
\begin{equation}
|\psi(z)\rangle=Se^{i g\Lambda z}S^{-1}|\psi(0)\rangle\equiv R|\psi(0)\rangle,
\label{R}
\end{equation}
with $R$ a (square) matrix of dimension $N$, whose elements are given by
\begin{equation*}
    R_{p,q}(z)=\frac{1}{\sqrt{2^{p+q}(p-1)!(q-1)!}}
\end{equation*}
\begin{equation}
\times\sum_{j=0}^{N-1}\frac{H_{p-1}\left(\displaystyle\frac{\lambda_j}{\sqrt{2}}\right)H_{q-1}\left(\displaystyle\frac{\lambda_j}{\sqrt{2}}\right)\exp(i g\lambda_jz) }{\displaystyle\sum_{k=0}^{N-1}\frac{1}{2^kk!}H_k^2\left(\frac{\lambda_j}{\sqrt{2}}\right)},
\label{Rpq}
\end{equation}
while $\Lambda$ is diagonal, its elements are $\left\{\lambda_j\right\}$.
By means of (\ref{R}) and (\ref{Rpq}) we get
\begin{equation}
    c_n(z)=\sum_{\ell=0}^{N-1}R_{n+1,\ell+1}(z) c_\ell(0),
\end{equation}
or equivalently
\begin{equation*}
c_n(z)=\frac{1}{\sqrt{2^nn!}}\sum_{\ell=0}^{N-1}\sum_{j=0}^{N-1}
    \frac{1}{\sqrt{2^\ell\ell!}}\end{equation*}
\begin{equation}
\times\frac{H_n\left(\displaystyle\frac{\lambda_j}{\sqrt{2}}\right)H_\ell\left(\displaystyle\frac{\lambda_j}{\sqrt{2}}\right) \exp(i g\lambda_jz)}{\displaystyle\sum_{k=0}^{N-1}\frac{1}{2^kk!}H_k^2\left(\frac{\lambda_j}{\sqrt{2}}\right)}c_\ell(0).
\end{equation}
\textit{Response to the impulse.} Again, when the initial condition $|\psi(0)\rangle$ is set as exciting just one waveguide, for example that at the $m$-th site, then $c_\ell(0)=\delta_{\ell m}$, where $\delta_{ij}$ stands for the the Kronecker delta symbol, then
\begin{equation}
c_n(z)=\frac{1}{\sqrt{2^{n+m}n!m!}}
\end{equation}
\begin{equation}
\times\sum_{j=0}^{N-1}
\frac{H_n\left(\displaystyle\frac{\lambda_j}{\sqrt{2}}\right)H_m\left(\displaystyle\frac{\lambda_j}{\sqrt{2}}\right)\exp(i g\lambda_jz)}{\displaystyle\sum_{k=0}^{N-1}\frac{1}{2^kk!}H_k^2\left(\frac{\lambda_j}{\sqrt{2}}\right)}.
\label{finitec}
\end{equation}
In turn, the corresponding solution in the non-Hermitian array is just 
\begin{equation}
d_n(z)=\frac{e^{\gamma n}}{\sqrt{2^{n+m}n!m!}}
\end{equation}
\begin{equation}
\times\sum_{j=0}^{N-1}
\frac{H_n\left(\displaystyle\frac{\lambda_j}{\sqrt{2}}\right)H_m\left(\displaystyle\frac{\lambda_j}{\sqrt{2}}\right)\exp(i g\lambda_jz)}{\displaystyle\sum_{k=0}^{N-1}\frac{1}{2^kk!}H_k^2\left(\frac{\lambda_j}{\sqrt{2}}\right)}.
\label{finited}
\end{equation}
In Figure \ref{fig.finite1} the intensities $|d_n(z)|^2$ are shown, as ruled by (\ref{finited}), for some specific values of the parameters. By comparison with the semi-infinite scenario, it is seen that in this case reflections appear in the two edge waveguides (see lower rows in both Figure \ref{fig.finite1} and Figure \ref{fig.semiinfinite1}). In turn, they can be tailored by means of the parameter $\gamma$ giving the non-Hermitian nature to the array. Such can be appreciated from Figure \ref{fig.finite1} (lower middle and lower right panels). 
\\ \\
\textit{Supermodes of the finite array. }
If $|\psi(0)\rangle$ is now chosen as $|\psi(0)\rangle =|\psi_j\rangle$, where $|\psi_j\rangle$ satisfies (\ref{eva}), stationary evolution is obtained:
\begin{equation}
   |\psi(z)\rangle=\exp\left(i g\lambda_j z\right)|\psi_j\rangle.
\label{super}
\end{equation} These are known as the supermodes of the optical lattice \cite{Yariv}.

\begin{figure}[h]
    \centering
    {\includegraphics[width=\linewidth]{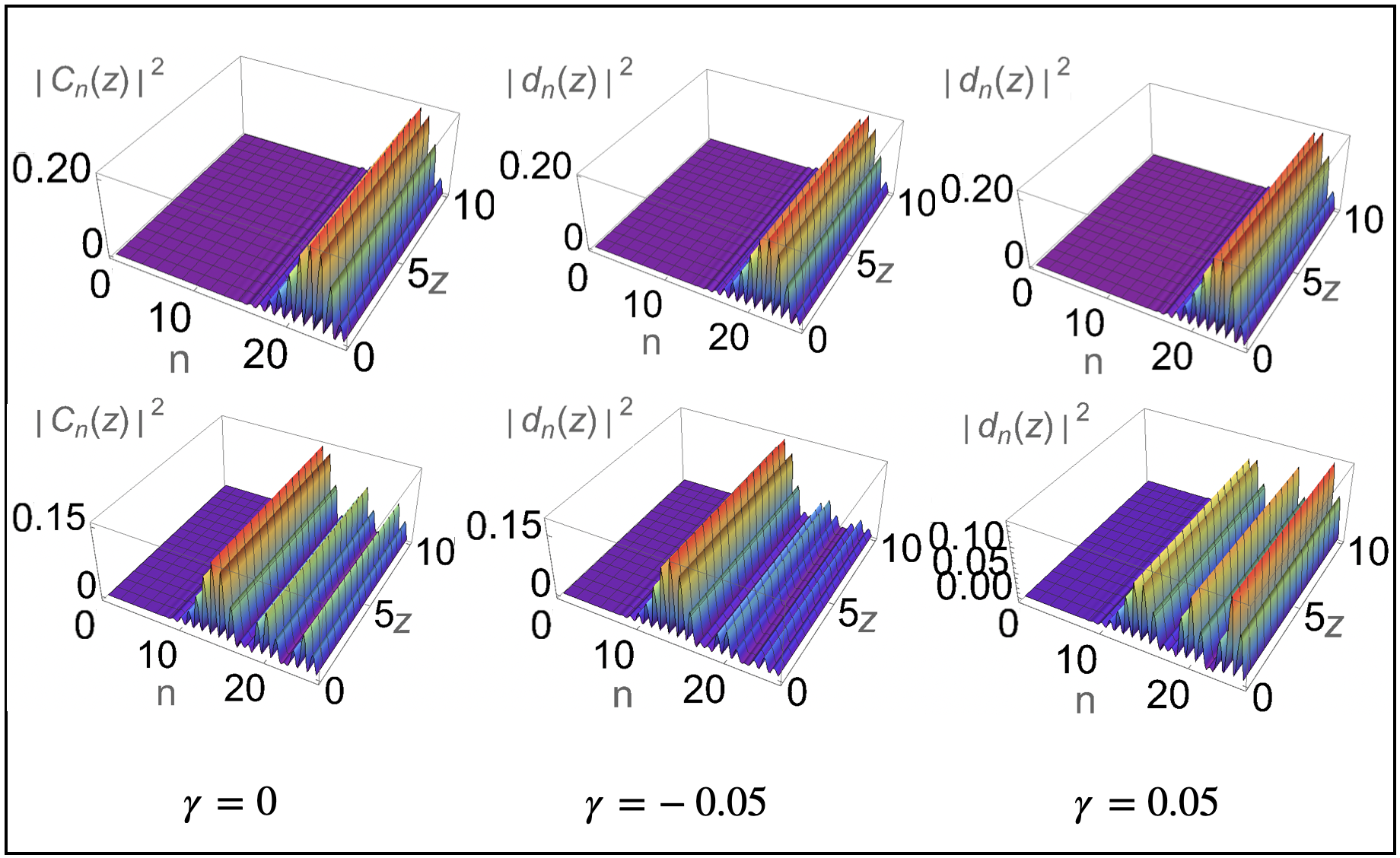}}
\caption{Stationary propagation of the supermodes $|\psi_j\rangle$ and $|\phi_j\rangle$, corresponding to an Hermitian and non-Hermtitian system of waveguides of the Glauber-Fock type. The array is formed of $N=25$ waveguides and the parameters coincide with those in Figure \ref{fig.finite1}. In the upper row the supermode $j=0$ is shown. The field is principally concentrated at the right of the array, at $n\sim 23$, in the Hermitian ($\gamma=0$) and non-Hermitian ($\gamma\neq 0$) regimes, for the indicated values of $\gamma$. In the lower row the supermode $j=4$ is shown. Once more, the Hermitian case in the lower left panel preserves the non-symmetric distribution of the field (with respect to the central waveguides). For the non-Hermitian system $\gamma=-0.05$ ($\gamma=0.05$) in the lower middle (lower right) panel, it can be appreciated a redistribution of field caused by the $\gamma\neq 0$.}
	\label{fig.finite2}
\end{figure}

In addition, the choice $|\phi(0)\rangle= e^{\gamma \hat n} |\psi_j\rangle$ gives the corresponding (not normalized) stationary evolution in the non-Hermitian system:
\begin{equation}
   |\phi(z)\rangle=\exp\left(i g\lambda_j z\right)e^{\gamma \hat n}|\psi_j\rangle. 
\label{super2}
\end{equation}
By replacing (\ref{super2}) into (\ref{nHermitian}), we obtain the eigenvalue problem for $\tilde M$
\begin{equation}
    \tilde M |\phi_j\rangle=\lambda_j |\phi_j\rangle,\qquad \tilde M=-\tilde H,
\end{equation}
where $|\phi_j\rangle=e^{\gamma \hat n}|\psi_j\rangle$. In figure \ref{fig.finite2} the supermodes of the Hermitian (left column) and non-Hermitian (middle and right columns) systems, $|\psi_j\rangle$ and $|\phi_j\rangle$, respectively, are presented for comparison. They were obtained numerically. The supermodes $j=0$ and $j=4$ are given, respectively, in the upper and lower rows. It is evident from Figure \ref{fig.finite2} that the supermodes are indeed stationary states of the electromagnetic field.
\section{SU(1,1) waveguide array}
\label{sec.su11}
In this section we consider a variation of the Glauber-Fock lattice considered in the previous sections \cite{Perez-Leija2011}. We consider a semi-infinite array in which the electromagnetic field $c_n(z)$ in the $n$-th waveguide satisfies
\begin{equation}
    i\dot c_n(z)+  g[f(n)c_{n-1}(z)+f(n+1)c_{n+1}(z)]=0,
    \label{su}
\end{equation}
where $f(n)=\sqrt{\displaystyle\frac{n+\xi n^2}{\xi}}$, $\xi\in\mathbb R$. Equation (\ref{su}) has a partner equation (\ref{quantum}), with $H$ given by
\begin{equation}
    H=- g(A+A^\dagger),
    \label{Hsu}
\end{equation}
\begin{equation}
   A=a\sqrt{\frac{1+\xi\hat n}{\xi}},
\end{equation}
\begin{equation}
A^\dagger=\sqrt{\frac{1+\xi\hat n}{\xi}}a^\dagger,
\end{equation}
with $a$ and $a^\dagger$ the annihilation and creation operators defined in Section \ref{sec.semi}. Also as in Section \ref{sec.semi}, (\ref{su}) and (\ref{quantum}) connect by means of (\ref{combination}). The corresponding commutation relations are
    \begin{equation}
    [A,A^\dagger]=2\hat n+\frac{1}{\xi}+1,
    \label{commutationsu1}
\end{equation}
    \begin{equation}
    [\hat n,A]=-A,
    \end{equation}
    \begin{equation}
    [\hat n,A^\dagger]=A^\dagger.
    \end{equation}

The solution of (\ref{quantum}), for $H$ as given in (\ref{Hsu}) is then
\begin{equation}
    |\psi(z)\rangle=e^{i g z(A+A^\dagger)}|\psi(0)\rangle.
    \label{solsu}
\end{equation}
The commutator (\ref{commutationsu1}) once more forbids to express the evolution operator $\exp[i g z(A+A^\dagger)]$ in (\ref{solsu}) in the factorized form (\ref{Glaub}). Nevertheless, by introducing the operator $A_0=\hat n+\displaystyle\frac{1}{2\xi}+\displaystyle\frac{1}{2}$, it is straightforward to get the commutation relations of the $SU(1,1)$ operator algebra for the operators $\left\{A_0, A, A^\dagger\right\}$. These are,
\begin{equation}
    [A,A^\dagger]=2A_0,\quad [A_0,A]=-A,\quad [A_0,A^\dagger]=A^\dagger.
\end{equation}
By proposing
\begin{equation}
|\psi(z)\rangle=e^{iu(z)A^\dagger}e^{v(z)A_0} e^{iw(z)A}|\psi(0)\rangle, 
\end{equation}
subject to the initial conditions $u(0)=v(0)=w(0)=0$, it is straightforward to obtain $u(z)=w(z)=\tanh  g z$, and $v(z)=\ln(\displaystyle\frac{1}{\cosh^2  g z})$. 
\\ \\
\textit{Response to the impulse.} For $|\psi(0)\rangle=|m\rangle$,
\begin{equation*}
    |\psi(z)\rangle=(\cosh  g z)^{-\displaystyle\frac{1+(2m+1)\xi}{\xi}}[f(m)]!
\end{equation*}
\begin{equation}
\times\sum_{j,k=0}^{m,\infty}\frac{\cosh^{2j} g z (iu)^{j+k}[f(m-j+k)]!}{j!k!([f(m-j)]!)^2}|m-j+k\rangle,
\end{equation}
where $[f(\ell)]!=f(\ell)f(\ell-1)\dots f(1)$, with $[f(0)]!=1$. Therefore the electric field in the $n$-th waveguide is given by
\begin{equation*}
   c_n(z)=(\cosh  g z)^{-\displaystyle\frac{1+(2m+1)\xi}{\xi}}(i\tanh g z)^{-m+n}
\end{equation*}
\begin{equation}
\times[f(m)]![f(n)]!\sum_{j=0}^m\frac{(-1)^j\sinh^{2j} g z}{j!(n+j-m)!([f(m-j)]!)^2},
\end{equation}
and the fields in the non-conservative system are simply
\begin{equation*}
   d_n(z)=e^{\gamma n}(\cosh  g z)^{-\displaystyle\frac{1+(2m+1)\xi}{\xi}}(i\tanh  g z)^{-m+n}
\end{equation*}
\begin{equation}
\times[f(m)]![f(n)]!\sum_{j=0}^m\frac{(-1)^j\sinh^{2j} g z}{j!(n+j-m)!([f(m-j)]!)^2}.
\label{dnsu11}
\end{equation}

\begin{figure}[h]
    \centering
    {\includegraphics[width=\linewidth]{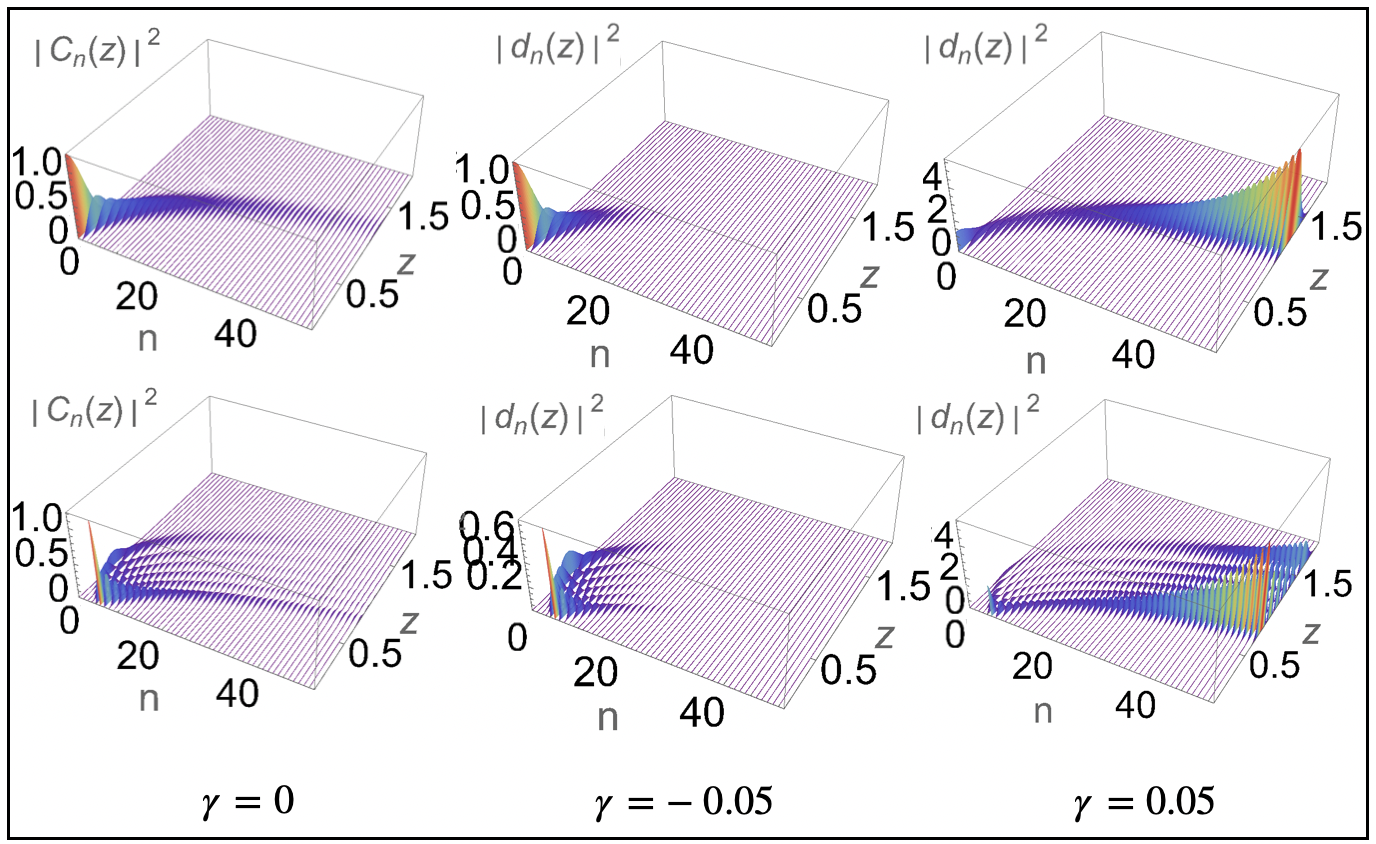}}
\caption{Non-Hermitian propagation of the intensity $|d_n(z)|^2$ in the semi-infinite Glauber-Fock optical lattice, as given by (\ref{dnsu11}), in a semi-infinite $SU(1,1)$ waveguide array when one single waveguide situated at $m$ is at $z=0$. In all the figures the longitudinal distance $z$ is measured in units of the coupling constant $g$, while the parameter $g= 1$ was set. Quite generally, a behavior very close to the one observed in Figure \ref{fig.semiinfinite1} can be appreciated. For the chosen value of $\xi$, smaller distance $z$ is reached, in comparison to the Glauber-Fock lattice in Figure \ref{fig.semiinfinite1}. Upper row: the edge waveguide $m=0$ is initially excited. Similar to the upper row in Figure \ref{fig.semiinfinite1}, the Hermitian $\gamma=0$ (upper left) and non-Hermitian cases $\gamma=-0.05$ (upper middle) and  $\gamma=0.05$ (upper right) are characterized, respectively, by no attenuated/amplified, attenuated and amplified transport (see the vertical scales) corresponding to closed (upper left) and open systems (upper middle and upper right). Lower row: the site $m=5$ is initially excited. For the Hermitian case $\gamma=0$ (lower left) there is no attenuation or amplification. In turn, for the non-Hermitian cases $\gamma=-0.05$ (lower middle) and $\gamma=0.05$ (lower right), the effect of the transformation (\ref{phipsi}) can be clearly appreciated, \textit{i.e.} an attenuation and amplification towards $n\to\infty$, respectively.}
	\label{fig.su11}
\end{figure}

Figure \ref{fig.su11} shows a comparison between the non-Hermitian transport, given by (\ref{dnsu11}) and the Hermitian one ($\gamma=0$), for specific values of the parameters. Quite generally, a behavior pretty close to the one shown in Figure \ref{fig.semiinfinite1} can be appreciated. For the chosen value of $\xi$, the major difference between the Glauber-Fock and the $SU(1,1)$ arrays (Figure \ref{fig.semiinfinite1} and Figure \ref{fig.su11}, respectively) is the distance $z$ reached.
\section{Driven Glauber-Fock lattice}
\label{sec.driven}
\begin{figure}[h]
    \centering
    {\includegraphics[width=\linewidth]{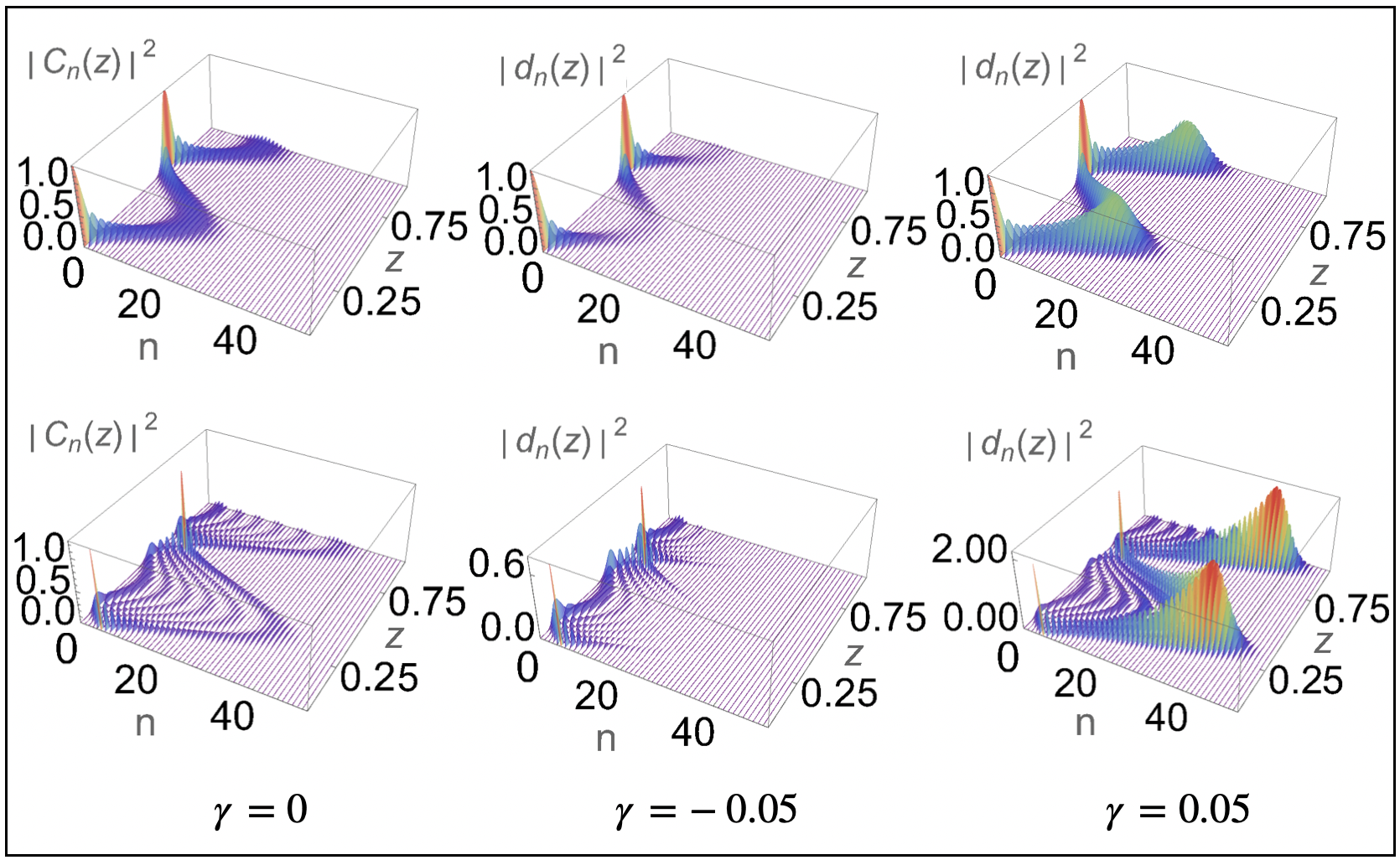}}
\caption{Non-Hermitian propagation of the intensity $|d_n(z)|^2$ in the semi-infinite Glauber-Fock optical lattice, as given by (\ref{dndriven}), for a semi-infinite \textit{driven} Glauber-Fock waveguide array when one single waveguide $m$ is excited at $z=0$. For all the figures $z$ is measured in units of $ g$, and the parameters $\omega=1$, $ g=2$ units have been chosen. Upper row: For $m=0$ a single maximum of the electromagnetic field can be seen, following a periodic trajectory along $z$. This is so due to the first term in both (\ref{drivenH}) and (\ref{nHwn}), and can be particularly appreciated in the Hermitian case $\gamma=0$ (upper left). For the non-Hermitian cases (upper middle and upper right) $\gamma=-0.05$ and $\gamma=0.05$ a subtle attenuation and amplification in the waveguides at the right of the array (at $n\to\infty$) can be noticed, according to (\ref{phipsi}). For $m=5$ (lower row) six maxima can be distinguished, following again a periodic propagation along $z$ (see for instance the Hermitian case $\gamma=0$ in the lower left panel). As before, the transformation (\ref{phipsi}) produces attenuation and amplification as $n\to\infty$ in the non-Hermitian cases $\gamma=-0.05$ and $\gamma=0.05$ shown in the lower middle and lower right panels, respectively. Such modulation can be of course be tailored by adjusting the non-Hermitian parameter $\gamma$.}
\label{fig.driven1}
\end{figure}

\begin{figure}[h]
    \centering
    {\includegraphics[width=\linewidth]{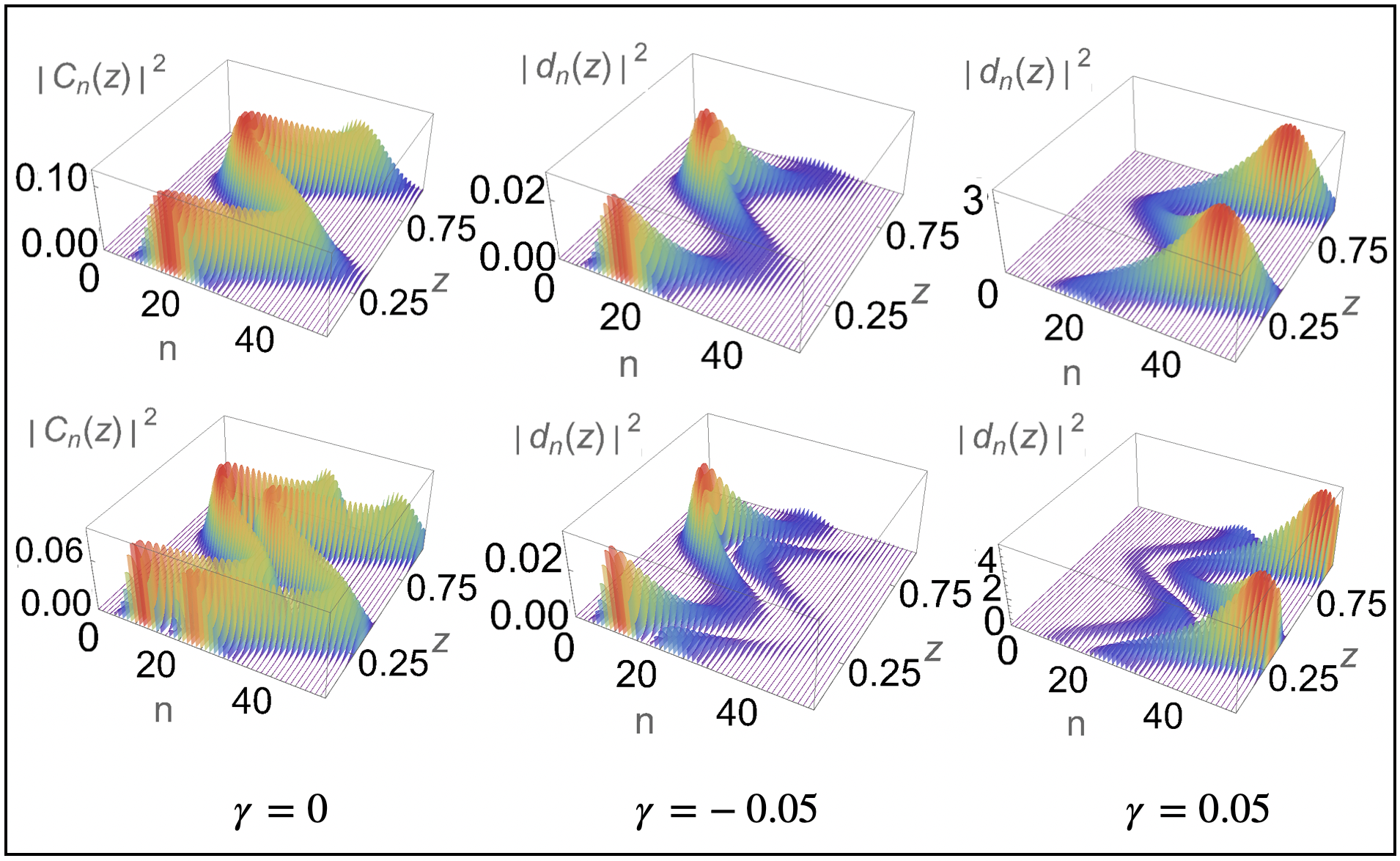}}
\caption{Non-Hermitian propagation of the intensity $|d_n(z)|^2$ in the semi-infinite Glauber-Fock optical lattice, as given by (\ref{csdriven}), for a semi-infinite \textit{driven} Glauber-Fock waveguide array when it is excited at $z=0$ with a distribution (a \textit{generalized coherent state}). The parameters are the same of Figure \ref{fig.driven1} and $\chi$ is chosen equal $4$. For $m=0$ (upper row) the distribution (Poissonian at $z=0$) has one maximum that propagates periodically in $z$, in a way \textit{somehow} close to the one shown in the upper row of Figure \ref{fig.driven1}. This can be particularly appreciated in the Hermitian case $\gamma=0$ (upper left panel). In the non-Hermitian cases $\gamma=-0.05$ and $\gamma=0.05$, a slight attenuation and amplification towards $n\to\infty$ can be observed (upper middle and right), once more similarly to the corresponding plots in Figure \ref{fig.driven1}. For $m=1$ (lower row), again a distribution is injected to the array at $z=0$. This time two lobes or maxima can be observed (see for instance the Hermitian case $\gamma=0$ in the lower left panel), following a periodic transport along $z$ as well. This, in turn, resembles the number of maxima in the lower row of Figure \ref{fig.driven1} as $m$ increases. On the other hand, in the non-Hermitian cases $\gamma\neq 0$ (lower middle and right) the relative heights of such lobes can be adjusted at will by changing the non-Hermitian parameter $\gamma$ and without affecting the periodic propagation. Naturally, this coherent states become the ones in Figure \ref{fig.semiinfinite2} in the limit $\omega\to 0$.}
\label{fig.driven2}
\end{figure}

A different modification of the Glauber-Fock lattice reviewed in Section \ref{sec.semi} is examined here. The following Hamiltonian is considered
\begin{equation}
    H=-\omega \hat n- g(a^\dagger+a),\qquad \omega,  g,\in\mathbb R,
\label{drivenH}
\end{equation}
with $a$, $a^\dagger$ and $\hat n$ as given in Section \ref{sec.semi}. Equation (\ref{quantum}), for the Hamiltonian (\ref{drivenH}), is related to the equation 
\begin{equation}
    i\dot c_n+\omega nc_n + g(\sqrt{n}c_{n-1}+\sqrt{n+1}c_{n+1})=0.
\label{Hwn}
\end{equation}
In turn we have equation (\ref{nHermitian}), with $\tilde H$ given by
\begin{equation}
   \tilde H=-\omega \hat n-(k_1a^\dagger+k_2a),
\label{nHwn}
\end{equation}
connected to
\begin{equation}
    i\dot d_n+\omega n d_n+k_1 \sqrt{n} d_{n-1}+k_2 \sqrt{n+1}d_{n+1}=0.
\end{equation}
By doing $|\psi(z)\rangle=D^\dagger(\displaystyle\frac{ g}{\omega})|w(z)\rangle$ in (\ref{quantum}), we arrive at the following equation for $|w(z)\rangle$
\begin{equation}
    i\frac{\partial |w(z)\rangle}{\partial z} = \bar H|w(z)\rangle,\qquad \bar H:=-\omega\hat n+\frac{ g^2}{\omega}.
\end{equation}
Therefore, 
\begin{equation}
|\psi(z)\rangle=D^\dagger\left(\frac{ g}{\omega}\right)e^{iz(\omega \hat n-\frac{ g^2}{\omega})}D\left(\frac{ g}{\omega}\right)|\psi(0)\rangle.
\label{driv}
\end{equation}
\textit{Response to the impulse}. If at $z=0$ only the $m$-th waveguide is excited, then $|\psi(0)\rangle=|m\rangle$ in (\ref{driv}). By using 
\begin{equation}
    e^{iz\omega \hat n}D(\xi)=D(\xi e^{iz\omega})e^{iz\omega\hat n}, \qquad \xi\in\mathbb C,
\end{equation}
and by defining $\Gamma=\Gamma( g,\omega,z):=\frac{ g}{\omega}(e^{iz\omega}-1)$, after some algebra we obtain
\begin{equation}
c_n(z)=e^{i\theta}e^{-\frac{1}{2}|\Gamma|^2}\Gamma^{n-m}\sqrt{\frac{m!}{n!}}L_m^{n-m}(|\Gamma |^2),
\end{equation}
where $\theta=\theta( g,\omega,z)=\frac{ g^2}{\omega^2}\sin(\omega z)-z\frac{ g^2}{\omega}+z\omega m$. Therefore the electric field in the non-conservative system is simply
\begin{equation}
d_n(z)=e^{\gamma n
}e^{i\theta}e^{-\frac{1}{2}|\Gamma|^2}\Gamma^{n-m}\sqrt{\frac{m!}{n!}}L_m^{n-m}(|\Gamma |^2).
\label{dndriven}
\end{equation}
Figure \ref{fig.driven1} shows the transport of the electromagnetic field as ruled by (\ref{dndriven}), for specific values of the parameters, as well as the comparison with the Hermitian case ($\gamma=0$). It can be observed that the $-\omega\hat n$ term in both (\ref{drivenH}) and (\ref{nHwn}) produces periodic propagation, unlike the one observed, for instance, in Figure \ref{fig.semiinfinite1}. Naturally, we recover the typical Glauber-Fock response in the limit $\omega\to 0$.
\\ \\
\textit{Coherent states as initial condition.} Now we chose $|\psi(0)\rangle$ in (\ref{driv}) as $|\psi(0)\rangle=|m,\chi\rangle$, as in Section \ref{sec.semi}. After some calculations, and by defining $\Delta=\Delta( g,\omega,\chi,z):=(\frac{ g}{\omega}+\chi)e^{iz\omega}-\frac{ g}{\omega}$, we obtain
\begin{equation}
    c_n(z)=e^{i\Theta}e^{-\frac{1}{2} |\Delta |^2}\Delta^{n-m}\sqrt{\frac{m!}{n!}}L_m^{n-m}(|\Delta |^2),
\end{equation}
with $\Theta=\Theta( g,\omega,\chi,z):=\frac{ g}{\omega}(\frac{ g}{\omega}+\chi_R)\sin(\omega z)-z\frac{ g^2}{\omega}+z\omega m-\frac{ g}{\omega}\chi_I(1-\cos(\omega z))$, and where $\chi_I$ denotes the imaginary part of $\chi$. Therefore, the electromagnetic field in the non-conservative system is given by
\begin{equation}
    d_n(z)=e^{\gamma n}e^{i\Theta}e^{-\frac{1}{2} |\Delta |^2}\Delta^{n-m}\sqrt{\frac{m!}{n!}}L_m^{n-m}(|\Delta |^2).
    \label{csdriven}
\end{equation}
Figure \ref{fig.driven2} shows the propagation of the electromagnetic field for some values of the parameters, according to (\ref{csdriven}). A comparison with the corresponding Hermitian case ($\gamma=0$) is shown as well. As in the case of the impulse (Figure \ref{fig.driven1}), it can be seen that the number of maxima (this time in the distribution) increases with $m$. Also similar to Figure \ref{fig.driven1}, periodic transport can be appreciated, which remains unaffected by the transformation (\ref{phipsi}). In turn, this last allows to tailor the relative heights of the corresponding lobes or maxima through the non-Hermitian parameter $\gamma$.
\section{Conclusions}
\label{sec.conclusions}
We have analyzed several systems associated with the conventional Glauber-Fock waveguide array in both the semi-infinite and infinite regimes. The propagation in the non-conservative (non-Hermitian) systems is simply related to the corresponding in the conservative (Hermitian) one through the transformation (\ref{phipsi}). Closed analytical solutions for different initial conditions have been given. 

Quite generally, the transformation (\ref{phipsi}) produces an attenuation or amplification towards $n\to\infty$ without changing the trajectory of the maxima of the electromagnetic field. Such can be tailored by manipulating the non-Hermitian parameter $\gamma$. In turn, this have a wide variety of applications. For instance, it can be used to analyze the response of non-conservative (non-Hermitian) systems by means of conservative (Hermitian) ones. Also, the model can be used to simulate imperfections in the couplings between waveguides, which are always assumed to be reciprocal (isotropic). The transformation can be used also as a protocol of communication and/or in order to encrypt optical information.

Besides, the current manuscript is intended to give a deeper insight into the understanding of non-Hermitian effects and the behavior of non-Hermitian systems. In particular in systems of the Glauber-Fock type. As mentioned, the transformation (\ref{phipsi}) can be regarded as a (non-unitary) supersymmetric (SUSY) transformation connecting the Hamiltonians $H$ and $\tilde H$.

\section*{Acknowledgments}
I. Bocanegra acknowledges CONAHCyT (M\'exico) for financial support through the postdoctoral fellowship 711878 and projects A1-S-24569 and CF19-304307. He is also grateful to IPN (M\'exico) for supplementary economical support through the project SIP20232237.


\begin{thebibliography}{1}

\bibitem{Christodoulides2003}
D.~N. Christodoulides, F.~Lederer and Y.~Silberberg, ``Discretizing light behaviour in linear and nonlinear waveguide lattices", Nature \textbf{424}, 817--823 (2003).
\bibitem{Yariv}
A.~Yariv and P.~Yeh, ``Photonics: optical electronics in modern communications", Oxford University Press: New York, 6th Ed. (2007).
\bibitem{Morandotti1999}
R.~Morandotti, U.~Peschel, J.~S. Aitchison, H.~S. Eisenberg and Y.~Silberberg, ``Experimental observation of linear and nonlinear optical Bloch oscillations", Phys. Rev. Lett. \textbf{83} (23), 4756--4759 (1999).
\bibitem{Eisenberg2000}
H.~S. Eisenberg, Y.~Silberberg,
R.~Morandotti and J.~S. Aitchison, ``Diffraction management", Phys. Rev. Lett. \textbf{85} (9) 1863--1866 (2000).
\bibitem{Pertsch2002}
T.~Pertsch, T.~Zentgraf, U.~Peschel, A.~Br\"auer and F.~Lederer, ``Anomalous refraction and diffraction in discrete optical systems", Phys. Rev. Lett. \textbf{88} (9), 093901-1--4 (2002).
\bibitem{Perets2008}
H.~B. Perets, Y.~Lahini, F.~Pozzi, M.~Sorel, R.~Morandotti and Y.~Silberberg, ``Realization of quantum walks with negligible decoherence in waveguide lattices", Phys. Rev. Lett. \textbf{100}, 170506-1--4 (2008).
\bibitem{Leon-Montiel2010}
R.~de J. Le\'on-Montiel and H.~M. Moya-Cessa, ``Modeling non-linear coherent states in fiber arrays", Int. J. Quant. Information \textbf{9}, 349--355 (2010).
\bibitem{Leon-Montiel2011}
R.~de J. Le\'on-Montiel, H.~M. Moya-Cessa and F.~Soto-Eguibar, ``Nonlinear coherent states for the Susskind-Glogower operators", Rev. Mex. F\'is. \textbf{57} (3), 133-147 (2011).
\bibitem{Perez-Leija2010}
A.~Perez-Leija, H.~M. Moya-Cessa, A.~Szameit and D.~N. Christodoulides, ``Glauber-Fock photonic lattices", Opt. Lett. \textbf{35} (14), 2409--2411 (2010).
\bibitem{Keil2011}
R.~Keil, A.~Perez-Leija, F.~Dreisow, M.~Heinrich, H.~Moya-Cessa, S.~Nolte, D.~N. Christodoulides and A.~Szameit, ``Classical analogue of displaced Fock states and quantum correlations in Glauber-Fock photonic lattices", Phys. Rev. Lett. \textbf{107}, 103601-1--5 (2011).
\bibitem{Villegas-Martinez2022}
B.~M. Villegas-Mart\'inez, H.~M. Moya-Cessa and F.~Soto-Eguibar, ``Modeling displaced squeezed number states in waveguide arrays", Physica A: Statistical Mechanics and its applications \textbf{68}, 128265 (2022).
\bibitem{Iwanow2005}
R.~Iwanow, D.~A. May-Arrioja, D.~N. Christodoulides and G.~I. Stegeman, ``Discrete Talbot effect in waveguide arrays", Phys. Rev Lett. \textbf{95}, 053902-1--4 (2005).
\bibitem{Rai2008}
A.~Rai, G.~S. Agarwal and J.~H. H. Perk, ``Transport and quantum walk of nonclassical light in coupled waveguides", Phys. Rev. A \textbf{78}, 042304-1--5 (2008).
\bibitem{Keil2012}
R.~Keil, A.~Perez-Leija, P.~Aleahmad, H.~Moya-Cessa, S.~Nolte, D.~N. Christodoulides and A.~Szameit, ``Observation of Bloch-like revivals in semi-infinite Glauber-Fock photonic lattices", Opt. Lett. \textbf{37} (18), 3801--3 (2012).
\bibitem{Perez-Leija2012}
A.~Perez-Leija, R.~Keil, A.~Szameit, A.~F. Abouraddy, H.~Moya-Cessa and D.~N. Christodoulides, ``Tailoring the correlation and anticorrelation behavior of path-entangled photons in Glauber-Fock oscillator lattices", Phys. Rev. A \textbf{85} 013848-1--5 (2012).
\bibitem{Szameit2007}
A.~Szameit, F.~Dreisow, H.~Hartung, S.~Nolte, A.~T\"unnermann and F.~Lederer, ``Quasi-incoherent propagation in waveguide arrays", App. Phys. Lett. \textbf{90}, 241113-1--3 (2007).
\bibitem{Bromberg2009}
Y.~Bromberg, Y.~Lahini, R.~Morandotti and Y.~Silberberg, ``Quantum and classical correlations in waveguide lattices", Phys. Rev. Lett. \textbf{102}, 253904-1--4 (2009).
\bibitem{Rodriguez-Lara2011}
B.~M. Rodr\'iguez-Lara, ``Exact dynamics of finite Glauber-Fock photonic lattices", Phys. Rev A \textbf{84}, 053845-1--6 (2011).
\bibitem{Longhi2010}
S.~Longhi, ``Photonic analog of Zitterbewegung in binary waveguide arrays", Opt. Lett. \textbf{35} (2), 235--237 (2010).
\bibitem{Dreisow2009}
F.~Dreisow, A.~Szameit, M.~Heinrich, T.~Pertsch, S.~Nolte and A.~T\"unnermann, ``Bloch-Zener oscillations in binary superlattices", Phys. Rev. Lett. \textbf{102}, 076802-1--4 (2009).
\bibitem{Heinrich2014}
M.~Heinrich, M.~A. Miri, S.~St\"utzer, R.~El-Ganainy, S.~Nolte, A.~Szameit and D.~N. Christodoulides, ``Supersymmetric mode converters", Nat. Commun. \textbf{5}, 1--7 (2014). 
\bibitem{Perez-Leija2011}
A.~Perez-Leija, H.~Moya-Cessa, F.~Soto-Eguibar, O.~Aguilar-Loreto and D.~N. Christodoulides, ``Classical analogues to quantum nonlinear coherent states in photonic lattices", Optics Communications \textbf{284} 1833--1836 (2011).
\bibitem{Perez-Leija2013}
A.~Perez-Leija, R.~Keil, H.~Moya-Cessa, A.~Szameit and D.~N. Christodoulides, ``Perfect transfer of path-entangled photons in $J_x$ photonic lattices", Phys. Rev. A \textbf{87}, 022303-1--7 (2013).
\bibitem{Perez-Leija2013b}
A.~Perez-Leija, J.~C. Hernandez-Herrejon H.~Moya-Cessa, A.~Szameit and D.~N. Christodoulides, ``Generating photon-encoded W states in multiport waveguide-array systems", Phys. Rev. A \textbf{87}, 013842-1--5 (2013).
\bibitem{Bender1998}
C.~M. Bender and S.~Boettcher, ``Real spectra in non-Hermitian Hamiltonians having PT symmetry", Phys. Rev. Lett. \textbf{80} (24), 5243--5246 (1998).
\bibitem{Bender1999}
C.~M. Bender, S.~Boettcher and P.~N. Meisinger, ``PT-symmetric quantum mechanics", J. Math. Phys. \textbf{40} (5), 2201--2229 (1999).
\bibitem{Mostafazadeh2003}
A.~Mostafazadeh, ``Exact PT symmetry is equivalent to Hermiticity", J. Phys. A: Math. Gen. \textbf{36}, 7081--7091 (2003).
\bibitem{Makris2008}
K.~G. Makris, R.~El-Ganainy, D.~N. Christodoulides and Z.~H. Musslimani, ``Beam dynamics in PT symmetric optical lattices", Phys. Rev. Lett. \textbf{100}, 103904-1--4 (2008).
\bibitem{Makris2010}
K.~G. Makris, R.~El-Ganainy, D.~N. Christodoulides and Z.~H. Musslimani, ``PT-symmetric optical lattices", Phys. Rev. A \textbf{81}, 063807-1--10 (2010).
\bibitem{Longhi2015}
S.~Longhi, D.~Gatti and G.~D. Valle, ``Robust light transport in non-Hermitian photonic lattices", Sci. Rep. \textbf{5}, 13376--13388 (2015).
\bibitem{Regensburger2012}
A.~Regensburger, C.~Bersch, M.~A. Miri, G.~Onishchukov, D.~N. Christodoulides and U.~Peschel, ``Parity-time synthetic photonic lattices", Nature \textbf{488}, 167--171 (2012).
\bibitem{Yuce2022}
C.~Yuce and H.~Ramezani, ``Diffraction-free beam propagation at the exceptional point of non-Hermitian Glauber Fock lattices", J. Opt. \textbf{24} 1--5 (2022).
\bibitem{Weidemann2020}
S.~Weidemann, M.~Kremer, T.~Helbig, T.~Hofmann, A.~Stegmaier, M.~Greiter, R.~Thomale and A. Szameit, ``Topological funneling of light", Science \textbf{368}, 311--314 (2020).
\bibitem{Liu2021}
S.~Liu, R.~Shao, S.~Ma, L.~Zhang, O.~You, H.~Wu, Y.~J. Xiang, T.~J. Cui and S.~Zhang, ``Non-Hermitian skin effect in a non-Hermitian electrical circuit", Research \textbf{2021}, 1--9 (2021).
\bibitem{Liu2022}
Y.~G. N. Liu, Y.~Wei, O.~Hemmatyar, G.~G. Pyrialakos, P.~S. Jung, D.~N. Christodoulides and M.~Khajavikhan, ``Complex skin modes in non-Hermitian coupled laser array", Light: Science and Applications \textbf{11} (1), 336--341 (2022).
\bibitem{Bocanegra2023ES}
I. Bocanegra, ``Non-Hermitian propagation in equally-spaced Hermitian waveguide arrays", https://doi.org/10.48550/arXiv.2307.06952 (2023).
\bibitem{Hatano}
N.~Hatano and D.~R. Nelson, ``Localization transitions in non-Hermitian quantum mechanics", Phys. Rev. Lett. \textbf{77} (3), 570--573 (1996).
\bibitem{Braulio}
B.~M. Villegas-Mart\'inez, F.~Soto-Eguibar, S.~A. Hojman, F.~A. Asenjo and H.~M. Moya-Cessa, ``Non-unitary transformation approach to PT dynamics", https://arxiv.org/pdf/2201.06536.pdf.
\bibitem{Ruter2010}
C.~E. R\"uter, K.~G. Makris, R.~El-Ganainy, D.~N. Christodoulides, M.~Segev and D.~Kip, ``Observation of parity-time symmetry in optics", Nat. Phys. \textbf{6}, 192--195 (2010).
\bibitem{Louisell}
W.~H. Louisell, ``Quantum statistical properties of radiation", Wiley: New York, (1990).
\bibitem{Cooper1995}
F.~Cooper, A.~Khare and U.~Sukhatme, ``Supersymmetry and quantum mechanics", Phys. Rep. \textbf{251} 267--385 (1995).

\bibitem{Bocanegra2022}
I.~Bocanegra and S.~Cruz y Cruz, ``Classes of balanced gain-and-loss waveguides as non-Hermitian potential hierarchies", Symmetry \textbf{14}, 432 (2022).

\bibitem{Bocanegra2023}
I.~Bocanegra, ``New families of complex hyperbolic-secant refractive-index profiles through the factorization method", J. Phys.: Conf. Ser. \textbf{2448}, 012015 (2023).

\bibitem{Abramowitz}
M.~Abramowitz and I.~A. Stegun, ``andbook of mathematical functions with formulas, graphs, and mathematical tables", Department of commerce: USA, 9th Ed. (1970).

\end{thebibliography}
\end{document}